\documentclass[11pt]{article}
\usepackage{epsf,amssymb,cite}
\newcommand{\ft}[2]{{\textstyle\frac{#1}{#2}}}
\newcommand{\hc}{{\rm h.c.}}
\def\twovec#1#2{\left(\begin{array}{c}
{#1}\\ {#2}\\
\end{array}
\right)}

\newsavebox{\uuunit}
\sbox{\uuunit}
    {\setlength{\unitlength}{0.825em}
     \begin{picture}(0.6,0.7)
        \thinlines
        \put(0,0){\line(1,0){0.5}}
        \put(0.15,0){\line(0,1){0.7}}
        \put(0.35,0){\line(0,1){0.8}}
       \multiput(0.3,0.8)(-0.04,-0.02){10}{\rule{0.5pt}{0.5pt}}
     \end {picture}}
\newcommand {\unity}{\mathord{\!\usebox{\uuunit}}}
\def\rmi{{\rm i}}
\def\rmd{{\rm d}}
\def\rme{{\rm e}}
\newcommand{\SU}{\mathop{\rm SU}}
\newcommand{\SO}{\mathop{\rm SO}}
\newcommand{\U}{\mathop{\rm {}U}}
\newcommand{\USp}{\mathop{\rm {}USp}}
\newcommand{\OSp}{\mathop{\rm {}OSp}}

\newcommand{\Sl}{\mathop{\rm {}S}\ell }

\newcommand{\alphaa}{a}
\newcommand{\betaa}{b}
\makeatletter
\@addtoreset{equation}{section}
\makeatother

\newcommand{\ILambda }{\Lambda }

\newcommand{\LX}{L}
\def\Re{\mathop{\rm Re}\nolimits}
\def\Im{\mathop{\rm Im}\nolimits}
\def\diag{\mathop{\rm diag}\nolimits}
\def\dim{\mathop{\rm dim}\nolimits}
 
\begin{document}
\begin{titlepage}
\begin{flushright}
  KUL-TF-02/03 \\
  ITF-2002/20 \\
  SPIN-2002/13\\
  hep-th/0205119
\end{flushright}
\vspace{.5cm}
\begin{center} {{\Large \bf Stable de Sitter Vacua from
$\mathcal{N}=2$ Supergravity \hskip 0.2cm $^\dagger$
}}\\
\vfill {\large Pietro Fr{\'e} $^1$, Mario Trigiante $^2$
and Antoine Van Proeyen $^3$} \\
\vfill
{ \sl
$^1$ Dipartimento di Fisica Teorica, Universit{\'a} di Torino, $\&$ INFN - Sezione di Torino\\
via P. Giuria 1, I-10125 Torino, Italy \\
\vskip 0.3cm $^2$ Spinoza Institute, Leuvenlaan 4, NL-3508 Utrecht, The
Netherlands\\
\vskip 0.3cm
$^3$ Instituut voor theoretische fysica, Katholieke Universiteit Leuven, \\
B-3001 Leuven, Belgium\\
}
\end{center}
\vfill
\begin{abstract}
We find extrema of the potential of matter couplings to $\mathcal{N}=2$
supergravity that define de Sitter vacua and no tachyonic modes. There
are three essential ingredients in our construction, namely non-Abelian
non-compact gaugings, de Roo--Wagemans rotation angles and
Fayet--Iliopoulos terms.
\end{abstract}
\vspace{2mm} \vfill \hrule width 3.cm
{\footnotesize
 $^ \dagger $ \hskip 0.1cm Work supported in part by the
European Community's Human Potential Programme under contract
HPRN-CT-2000-00131 Quantum Spacetime and  by the European Community Marie
Curie Fellowship under contract  HPMF-CT-2001-01276.}
\end{titlepage}
\newpage
\section{Introduction}
Recent cosmological observations lead to the conclusion that the
cosmological constant is positive and give a confirmation of the idea of
inflationary scenarios. Then, if string theory has to be able to provide
realistic models for cosmology, it should admit de Sitter vacua. The de
Sitter spaces are not as natural as anti-de Sitter ones in the context of
supersymmetric theories. This fact can clearly be seen from algebraic
considerations, and is illustrated in table~\ref{tbl:dSAdS}.
\begin{table}[ht]
  \caption{\it Superalgebras with bosonic subalgebra a direct product of (anti) de Sitter
  algebra and R-symmetry.}\label{tbl:dSAdS}
\begin{center}
\begin{tabular}{|c|l|rl|}\hline\hline
{\textbf AdS}& superalgebra&\multicolumn{2}{c|}{R-symmetry}\\ \hline
 $D=4$&$ \OSp(N|4) $ &&$ \SO(N) $\\   \hline
$D=5$&$ \SU(2,2|N)  $ &$N\neq 4:$&$  \SU(N)\times \U(1)$  \\
   &               &$N= 4:$& $  \SU(4)$  \\  \hline
$D=6$& F$^2(4)       $ &&$ \SU(2) $  \\       \hline
$D=7$&$ \OSp(6,2|N) $ &$N $ even:&$  \USp(N)$ \\
\hline \hline
{\textbf dS}  & superalgebra & \multicolumn{2}{c|}{R-symmetry} \\
\hline
$D=4$  & $\OSp(m^*|2,2)$ & $m=2$ & $\SO(1,1)$ \\
  &  & $m=4$ & $\SU(1,1)\times  \SU(2)$ \\
  &  & $m=6$ & $\SU(3,1)$ \\
  &  & $m=8$ & $\SO(6,2)$ \\
\hline
$D=5$  & $\SU^*(4|2n)$ & $n=1$ & $\SO(1,1)\times \SU(2)$ \\
  &  & $n=2$ & $\SO(5,1)$ \\
\hline
$D=6$  & F${}^1(4)$ &  & $\SU(2)$ \\
\hline \hline
\end{tabular}
\end{center}
\end{table}
The de Sitter
superalgebras~\cite{Lukierski:1985it,Lukierski:1984,Pilch:1985aw} have
typically a non-compact R-symmetry subalgebra\footnote{We mention here
the superalgebras that are of Nahm's type~\cite{Nahm:1978tg}, i.e., where
the bosonic subgroup is a direct product of the de Sitter algebra and
another simple group, called R-symmetry. Within this class, the $D=6$
case has a compact R-symmetry group. It has recently been shown
in~\cite{D'Auria:2002fh} that there are also ghosts in this case. More
general de Sitter superalgebras have been classified
in~\cite{D'Auria:2000ec,Ferrara:2001dn}.}, which leads to non-definite
signs in the kinetic terms, and hence leads to the existence of ghosts.
Therefore, de Sitter vacua can occur in physical supersymmetric theories
only in a phase where supersymmetry is completely broken. This might even
be welcome in view of the fact that supersymmetry breaking is anyhow
necessary to make contact with reality.

Furthermore, it has been mentioned that de Sitter vacua are difficult to
construct from higher dimensions~\cite{Gibbons:1984kp,Maldacena:2000mw},
although this may not be completely excluded. In any case, de Sitter
vacua have been found in 4-dimensional higher ${\cal N}$ supergravity
models~\cite{Gates:1983ct,deWit:1984pk,Hull:1985rt,Hull:1985ea,Cremmer:1985hj,Hull:1985wa,Hull:1988jw}.
In most cases, this was obtained by considering supergravities with a
gauged non-compact group
(see~\cite{Hull:1984vg,Hull:1984yy,Hull:1985rt,Cordaro:1998tx} for
$\mathcal{N}=8$, \cite{deRoo:1985jh} for $\mathcal{N}=4$,
\cite{Castellani:1986ka} for $\mathcal{N}=3$ and \cite{deWit:1984xe} for
$\mathcal{N}=2$, and also the new possibilities for $\mathcal{N}=8$ that
were recently found~\cite{Andrianopoli:2002mf,Hull:2002cv}). Such
solutions have been reconsidered
recently~\cite{Hull:2001ii,Hull:2001yg,Townsend:2001ea,Kallosh:2001gr,Gibbons:2001wy}
However, it has been mentioned that such vacua have tachyons, and even
that the negative masses of these tachyons often have a fixed ratio to
the cosmological constant~\cite{Kallosh:2001gr}. Indeed, normalizing the
scalars in the Lagrangian so that (for real scalars)
\begin{equation}
  {\cal L}=\ft12 e\,\partial _\mu \phi \partial ^\mu \phi - e\,V(\phi
  )\,,
 \label{normalScalars}
\end{equation}
many examples were found where at least one of the scalars has
\begin{equation}
  \frac{\partial }{\partial \phi } \frac{\partial }{\partial \phi } V= -2
  V\,.
 \label{eigenvaluem2}
\end{equation}
Therefore, the question still remained whether there are stable de Sitter
vacua in $\mathcal{N}\geq 2$ supergravity\footnote{The potentials for
$\mathcal{N}\geq 2$ are determined from gauging, while the one in
$\mathcal{N}=1$ may originate from an ad-hoc superpotential, with no
relation to higher dimensions or superstrings.}, and whether, in the case
of an affirmative answer, they could be lifted to full fledged string
theory. In the present paper, we give a positive answer to the first
question for $\mathcal{N}=2$, and some preliminary arguments why the
second question may also be answered affirmatively. Rigid $\mathcal{N}=2$
theories have already been used to construct inflation
scenarios~\cite{Kallosh:2001tm}. Difficulties to generalize this
construction to supergravities were identified. We thus make a first step
to overcome these problems.

In section~\ref{ss:flavors}, we give the ingredients that will turn out
to be necessary in the construction of $\mathcal{N}=2$ supergravities
with stable de Sitter vacua. Three specific models are then discussed in
section~\ref{ss:3models}, and the masses in their corresponding de Sitter
vacua are studied. In the first two models we find only positive mass
fields. In the third model, which has non-trivial hypermultiplets, there
are no negative masses, but there is a valley in the potential
corresponding to zero-mass fields. In section~\ref{ss:summOutlook}, we
discuss further steps that can be considered in order to lift our stable
de Sitter vacua first to $\mathcal{N}=4$ theories and later to the field
theory limit of superstrings. A first appendix gives explicit expressions
for the geometry items appearing in the quaternionic-K{\"a}hler manifold that
we consider. A second appendix gives a table with indices that are used
throughout the paper.

\section{Three ingredients}\label{ss:flavors}
In this section, we show that in $\mathcal{N}=2$ supergravity we can
obtain stable de Sitter vacua if we introduce three ingredients that turn
out to be all equally necessary. The framework is provided by the
coupling of vector multiplets based on the following choice for the
special K{\"a}hler manifold:
\begin{eqnarray}
\mathcal{SK}_n & = & \mathcal{ST}[2,n] \, \equiv \,
\frac{\mathrm{SU(1,1)}}{\mathrm{U(1)}} \, \times \,
\frac{\mathrm{SO(2,n)}}{\mathrm{SO(2)\times SO(n)}}\,. \label{spiffero}
\end{eqnarray}
The three essential ingredients are:
\begin{enumerate}
  \item \textbf{Non compact gauging}, namely the gauge group $\mathcal{G}_{\rm gauge}$ is a product of
  several factors and it involves the \textit{non-compact simple
  factor}  $\mathrm{SO(2,1)}$ times
several other compact factors, including $\mathrm{SO(3)}$ or
$\mathrm{U(1)}$ factors.
  \item \textbf{de Roo--Wagemans symplectic angles} that magnetically rotate
one gauge group with respect to another.
  \item \textbf{Fayet--Iliopoulos (FI)
terms} that are possible for either $\mathrm{SO(3)}$ or $\mathrm{U(1)}$
factors.
\end{enumerate}
In this section, we present the construction of $\mathcal{N}=2$
supergravity models involving the three ingredients listed above.
\subsection{Special geometry items}
The  models we will consider fall into the general framework of
matter-coupled supergravity that was extensively discussed in the
literature using both superconformal tensor calculus methods
\cite{deWit:1984rz,deWit:1984pk,deWit:1985px} and the more direct
geometric methods based on the rheonomic
approach~\cite{Castellani:1991b,Andrianopoli:1997cm}. Here we follow the
notations and conventions of \cite{Andrianopoli:1997cm} where the most
general form of an $\mathcal{N}=2$ supergravity action was given for an
arbitrary choice of the special K{\"a}hler manifold of the vector multiplets,
of the quaternionic manifold for hypermultiplets and with general
permissible gauging.

In order to construct the explicit form of our supergravity model, we
need to begin with the symplectic sections of special geometry. Following
the notations of \cite{Andrianopoli:1997cm}, we write the holomorphic
section as
\begin{equation}
  \Omega= \left( \begin{array}{c}
  X^\Lambda \\
  F_\Sigma
\end{array}\right),
\label{Omegabig}
\end{equation}
where
\begin{eqnarray}
X^\Lambda(S,y) & = & \left( \begin{array}{c}
  \ft 1 2 \, \left( 1+y^2\right)  \\
  \ft 12 \, {\rm i} \, (1-y^2) \\
  y^\alphaa
\end{array}\right)  \qquad ;\quad \alphaa=1,\dots , n\,, \quad \nonumber\\
F_\Lambda(S,y) & = & \left( \begin{array}{c}
  \ft 1 2 \, S \,  \left( 1+y^2\right)   \\
  \ft 12 \, {\rm i} \, S \,  (1-y^2)  \\
  - S \, y^\alphaa
\end{array}\right) \quad ;\quad y^2 = \sum_{\alphaa=1}^{n} (y^\alphaa)^2\,.
\label{symsecso21}
\end{eqnarray}
In the above equations, the complex fields $y^\alphaa$ are the
Calabi--Vesentini coordinates for the homogeneous manifold
$\frac{\mathrm{SO(2,n)}}{\mathrm{SO(2)\times SO(n)}}$, while the complex
field $S$ parametrizes the homogeneous space
$\frac{\mathrm{SU(1,1)}}{\mathrm{U(1)}}$ which is identified with the
complex lower half-plane. Indeed, the positivity domain of the Lagrangian
we are going to construct, implies
\begin{equation}
  \mbox{Im} \, S \, < \, 0\,.
\label{positdom}
\end{equation}
The  K{\"a}hler potential is, by definition, identified as
\begin{equation}
{\cal K}\,  = \,  -\mbox{log}\left ({\rm i}\langle \Omega \,
 \vert \, \bar \Omega
\rangle \right )\, =\, -\mbox{log}\left [ {\rm i} \left ({\bar X}^\Lambda
F_\Lambda - {\bar F}_\Sigma X^\Sigma \right ) \right ]. \label{specpot}
\end{equation}
The K{\"a}hler potential and metric associated with the above geometry are
\begin{eqnarray}
\mathcal{K} & = & \mathcal{K}_1 + \mathcal{K}_2 \,,\nonumber\\
\mathcal{K}_1  & = & -\log \, \left[ {\rm i} \, \left(
S-\overline{S}\right) \right],\qquad
 \mathcal{K}_2 = -\log \left[ \ft 1 2
\, \left( 1-2\overline{y}^\alphaa \, y^\alphaa + | y^\alphaa y^\alphaa|^2
\right) \right],
\nonumber\\
g_{S\overline{S}}&= & \frac{1}{(2\Im S)^2}\,,\qquad \,\qquad \, \quad
g_{\alphaa \bar \betaa}= \,\frac{\partial }{\partial
y^\alphaa}\,\frac{\partial }{\partial \bar {y}^\betaa}\, \mathcal{K}_2\,.
\label{kalermetr}
\end{eqnarray}
The covariantly holomorphic section is then defined by the general
formula
\begin{equation}
V \, = \, \twovec{L^{\Lambda}}{M_\Sigma} \, \equiv \, e^{{\cal
K}/2}\Omega \,= \, e^{{\cal K}/2} \twovec{X^{\Lambda}}{F_\Sigma}\,,
\label{covholsec}
\end{equation}
and satisfies the constraint
\begin{equation}
1 \, = \,  {\rm i}\langle V  \,
 \vert \, \bar V
\rangle  \, = \,   {\rm i} \left ({\bar L}^\Lambda M_\Lambda - {\bar
M}_\Sigma L^\Sigma \right )\,. \label{specpotuno}
\end{equation}
\subsection{Quaternionic-K{\"a}hler geometry items} \label{ss:QK}
The quaternionic-K{\"a}hler manifold with coordinates $q^u$, (and $u=1,\ldots
,4\dim QK$) has a metric build from vielbein 1-forms $V^{mt}=V^{mt}_u\rmd
q^u$, with $t=1,\ldots ,\dim QK$ and $m=1,\ldots ,4$:
\begin{equation}
  h_{uv}\rmd q^u \rmd q^v= \frac12\sum_{m,t} V^{mt}V^{mt}\,.
 \label{quatmetric}
\end{equation}
These vielbeins and their inverse $V_{mt}^u$ lead to the complex
structures ($x=1,2,3$)
\begin{equation}
  J^x{}_u{}^v=V_u^{mt}\mathbb{J}^x{}_m{}^n V_{nt}^v\,,
 \label{defhypercomplex}
\end{equation}
where
\begin{equation}
\mathbb{J}^1  =  \left( \matrix{ 0 & 1 & 0 & 0 \cr
    -1 & 0 & 0 & 0 \cr 0 & 0 & 0 & 1 \cr
   0 & 0 &
    -1 & 0  \cr  } \right)\,,  \qquad
\mathbb{J}^2  =  \left( \matrix{ 0 & 0 &
    -1 & 0 \cr 0 & 0 & 0 & 1 \cr 1 & 0 &
   0 & 0 \cr 0 &
    -1 & 0 & 0  \cr   }\right)\,,\qquad
\mathbb{J}^3  =  \left(\matrix{
   0 & 0 & 0 & 1  \cr 0 & 0 & 1 & 0  \cr
   0 & -1 & 0 & 0\cr
    -1 & 0 & 0 & 0  \cr   } \right),
\label{bfjx}
\end{equation}
such that the complex structures fulfil the  quaternionic algebra $J^x \,
J^y = -{\bf 1}_{4r} \, \delta^{xy} + \epsilon ^{xyz} \, J^z$.

The manifold has $\SU(2)$ curvature
\begin{equation}
  \Omega ^x=\Omega ^x_{uv}\rmd q^u\wedge \rmd q^v=\rmd \omega ^x+\ft12\varepsilon ^{xyz}
  \omega ^y\wedge \omega ^z=-\ft12 V^{mt}\wedge J^x{}_{mn}V^{nt}\,.
 \label{SU2curv}
\end{equation}

Triholomorphic Killing vectors $k_\Lambda ^u$ are related to moment maps
${\cal P}_\Lambda ^x$ by
\begin{equation}
  2k_\Lambda ^u\Omega ^x_{uv}= \nabla _v{\cal P}^x_\Lambda =\partial _v{\cal
  P}^x_\Lambda+\varepsilon ^{xyz}\omega _v^y{\cal P}^z_\Lambda\,.
 \label{introMomentMap}
\end{equation}
They should further satisfy a relation (equivariance condition)
\begin{equation}
  2k_\Lambda^u k_\Sigma ^v \Omega _{uv}^x-\varepsilon ^{xyz}{\cal P}_\Lambda
  ^y {\cal P}_\Sigma^z = f^\Delta {}_{\Lambda \Sigma }{\cal P}_\Delta ^x\,,
 \label{equivariance}
\end{equation}
where the structure constants are defined by
\begin{equation}
  k_\Lambda ^u\partial _u k_\Sigma^v - k_\Sigma  ^u\partial _u k_\Lambda
  ^v= -f^\Delta {}_{\Lambda \Sigma }k_\Delta^v\,.
 \label{defstructurec}
\end{equation}

\subsection{Description of the 3 ingredients}
\paragraph{Non-compact gauge groups.} In all previous
examples displaying de Sitter vacua, non-compact gauge groups were used.
In $\mathcal{N}=2$ tensor calculus, the appearance of just one
compensating multiplet leads to the non-compact factor gauge group
$\SO(2,1)$~\cite{deWit:1984xe}. In order to gauge non-Abelian groups
$\mathcal{G}_{\rm gauge}$, the special K{\"a}hler manifold of vector
multiplets must be a homogeneous space $\mathcal{G}/\mathcal{H}$, such
that $\mathcal{G}_{\rm gauge}\subset\mathcal{G}$, since the gauge
transformations must be continuous isometries of the scalar manifold.
Furthermore, for consistency, if we call $R$ the symplectic
representation of $\mathcal{G}$, to which the field strengths and their
magnetic duals are assigned, then, under the reduction to
$\mathcal{G}_{\rm gauge}$ we must have $R\stackrel{\mathcal{G}_{\rm
gauge}}{\longrightarrow }{\rm adj}+{\rm adj}$. Going through the list of
symmetric special manifolds~\cite{Cremmer:1985hc}, and especially their
symplectic embeddings (see, e.g., table 2 of~\cite{Fre:2001jd}), we see
that the only solution is the choice of the model~(\ref{spiffero}). In
this case, the electric and magnetic field strengths are in the doublet
representation of $\Sl(2,\mathbb{R})\sim\SU(1,1)$ and in the $n+2$ vector
representation of $\SO(2,n)$. For any compact group $G_{\rm compact}$ of
dimension $n-1$, the group $\SO(2,1)\times G_{\rm compact}$ is naturally
embedded in $\SO(2,n)$ in such a way that the vector $n+2=3+{\rm
adj}$(compact).

Hence, using $\mathcal{ST}\left[2,n\right]$, we can gauge a group of the
following type
\begin{eqnarray}
\mathcal{G}_{\rm gauge}&=& \mathrm{SO(2,1)}\times \mathrm{G}_1\times\dots\times \mathrm{G}_r\,,\nonumber\\
{\rm dim}(\mathrm{G}_k)&=& d_k \,;\,\qquad k=1,\dots,r\,, \label{gchoice}
\end{eqnarray}
where $\mathrm{G}_k$  are compact factors that can, in particular, be
$\mathrm{U(1)}$ or $\mathrm{SO(3)}$ factors. The condition on the
dimensions $d_k$ is obviously $ \sum_{k=1}^{r}\, d_k = n-1$. This is the
first essential ingredient in our identification of theories admitting
stable de Sitter vacua. We have introduced a gauge group with non-compact
generators.  For the $\mathrm{SO(2,1)}$ Lie algebra we use the following
normalization
\begin{eqnarray}
\left[  T_{x} \, , \, T_{y}\right]  & = & e_0 \, \varepsilon_{xyz} \,
\eta^{zw} \, T_{w}\,, \qquad
 x , y , \dots = 1,2,3\,,\nonumber\\
\left[  T_1 \, , \, T_2\right] & = & -T_3 \quad ; \quad \left[  T_1 \, ,
\, T_3\right]  =  -T_2  \,,\qquad  \left[  T_2 \, , \, T_3\right]  =
T_1\,, \label{algnormal}
\end{eqnarray}
where $e_0$ denotes the coupling constant of this group. Calling
collectively $t_\Lambda$ (with $ \Lambda=1,2, \dots, n+2$) the generators
of the gauge group (\ref{gchoice}) the structure constants of the gauge
Lie algebra are defined as follows:
\begin{equation}
  \left[ t_\Lambda \, , \, t_\Sigma\right] =
  f_{\Lambda\Sigma}^{\phantom{\Lambda\Sigma}\Delta} \, t_\Delta\,,
\label{gaugalge}
\end{equation}
and the symplectic embedding of the \textbf{adjoint representation} of
$\mathcal{G}_{\rm gauge}$ into the \textbf{fundamental representation} of
the symplectic group $\mathrm{Sp(2n+4,\mathbb{R})}$ is realized by
\begin{eqnarray}
 \mathcal{G}_{\rm gauge}  \ni \, t_\Lambda \, \hookrightarrow \,
  T_\Lambda & =& \left( \begin{array}{cc}
     t_\Lambda & 0 \\
    0 & -t^T_\Lambda    \
  \end{array} \right)\, \in  \, \mathrm{Sp(2n+4,\mathbb{R})}
 \,,\nonumber\\[2mm]
\left(  t_\Lambda\right) ^{\Sigma}_{\phantom{\Sigma}\Gamma} &=&f
_{\Lambda\Gamma}^{\phantom{\Lambda\Gamma}\Sigma}\,. \label{Tlamdefi}
\end{eqnarray}
Using (\ref{Tlamdefi}) we can write the \textit{real prepotentials} for
the Killing vectors describing the infinitesimal action
 of the gauge group on the scalar fields. We set
\begin{equation}
  P^0_\Lambda= \exp \left( \mathcal{K}\right)  < \overline{\Omega} \,|\,
  T_\Lambda \, \Omega >   \,,
\label{prepotente}
\end{equation}
and we have:
\begin{equation}
 \delta z^\alpha = \epsilon ^\Lambda k^\alpha_\Lambda (z)  \quad ; \quad
 k^\alpha_\Lambda (z)={\rm
 i} \, g^{\alpha \bar \beta} \, \partial _{\bar \beta} \, P^0_\Lambda\,,
\label{tuttivari}
\end{equation}
where $z^\alpha=\{S,y^0, \vec{y}\}$ denotes the entire set of all $n+1$
scalar fields. Applying (\ref{prepotente}) and (\ref{tuttivari}) to the
case of the $\mathrm{SO(2,1)}$ Lie algebra, we obtain the following
result for the Killing vectors:
\begin{eqnarray}
\vec{k}_1 & = & e_0 \, \left[ -{\rm i} \,\ft{1}{2} \,\left( 1 +\left(
y^0\right) ^2 -\left(\vec{ y}\right) ^2 \right) \,
\partial _0 -{\rm i} y^0 \,
\vec{y} \, \cdot \, \vec{\partial}\right] , \nonumber\\
\vec{k}_2 & = & e_0\, \left[ \,\ft{1}{2} \,\left( 1 -\left( y^0\right) ^2
+\left(\vec{ y}\right) ^2 \right) \, \partial _0 - y^0 \,
\vec{y} \, \cdot \, \vec{\partial}\right] , \nonumber\\
\vec{k}_3  & = & e_0 \left[ \,{\rm i} \, y^0 \, \partial _0 + {\rm i} \,
\vec{y} \, \cdot \, \vec{\partial}\right ] .\label{so21kil}
\end{eqnarray}
Note also that  the formula (\ref{prepotente}) for the Killing vector
prepotential is symplectic invariant, so that any symplectic rotation of
the section $\Omega$ does not affect the form of the Killing vector
fields.
\par
In the case where the compact part of the gauge group is just
$\mathrm{SO(3)}$, with Lie algebra normalized as follows:
\begin{equation}
  \left[  T_{x+3} \, , \, T_{y+3}\right]   =
  e_1 \,\varepsilon_{xyz} \, \,
  T_{z+3}
\quad ; \quad x,y,z , \dots = 1,2,3\,, \label{so3}
\end{equation}
and with $e_1$ denoting the associated  coupling constant, the Killing
vectors corresponding to these gauge group generators are
\begin{equation}
  \vec{k}_{x+3 } =e_1 \, \varepsilon _{xzw} \,
  y^z \, \frac{\partial}{\partial y^w}  \,, \qquad x,\,z,\,w
  = 1,2,3\,.
\label{so3killi}
\end{equation}
\paragraph{de Roo -- Wagemans angles.}
The second essential ingredient  is the introduction of \textit{de
Roo--Wagemans angles}, which were introduced in ${\cal N}=4$
supergravity\footnote{We learned from S.J. Gates, Jr. that in the earlier
paper~\cite{Gates:1984ha} occur parameters $\varphi$ and $\theta_0$ in
(3.1.1-2) and (3.2.1), respectively, which have the same effect.}
in~\cite{deRoo:1985jh,Wagemans:1990mv}, and which parameterize a rotation
of the relative embeddings of the $\mathrm{G_k}$ groups inside
$\mathrm{Sp(2({\bf n}+2),\mathbb{R})}$. These parameters are introduced
through a symplectic \textit{non-perturbative rotation} performed on the
holomorphic section of the manifold prior to gauging. Different choices
of the angles yield different gauged models with different physics. The
de Roo--Wagemans rotation matrix has the following form:
\begin{eqnarray}
\mathcal{R}&=&\left(\matrix{A & B\cr -B & A}\right)\,,\nonumber\\
 A&=& \left(\matrix{\unity_3 &0 &\dots & 0\cr
                         0  & \cos{(\theta_1)}\,\unity_{d_1} & & 0 \cr \vdots & & \ddots & \vdots\cr 0 &
                         0&\dots
                         &
                           \cos{(\theta_r)}\,\unity_{d_r}}\right),\nonumber\\
 B&=&\left(\matrix{0 &0 &\dots & 0\cr
                         0  & \sin{(\theta_1)}\,\unity_{d_1} & & 0 \cr \vdots & & \ddots & \vdots\cr 0 &
                         0&\dots
                         &
                           \sin{(\theta_r)}\,\unity_{d_r}}\right),
                           \label{dRWagangles}
\end{eqnarray}
the blocks being determined by the choice of the gauge group in
(\ref{gchoice}). The symplectic section is rotated in the following way:
\begin{eqnarray}
\Omega & \rightarrow & \Omega_R \, \equiv \, \mathcal{R}\cdot \Omega\,.
\end{eqnarray}
The K{\"a}hler potential is clearly left invariant by the above
transformation. In all calculations of the scalar potential, we have to
use the symplectic section $\Omega_R$ rather than $\Omega$ and $V_R
\equiv \exp \left[\mathcal{ K}\right] \, \Omega_R$ rather than $V$.
\paragraph{Fayet--Iliopoulos terms.}
Finally, the last ingredient we should introduce is the option of
including also Fayet--Iliopoulos terms for the $\mathrm{U(1)}$ or the
$\mathrm{SO(3)}$ factors that can appear in the compact part of the gauge
group. This possibility is of crucial importance at the level of our
analysis and is motivated by the following argument. In the absence of
quaternionic scalars, the equivariance condition~(\ref{equivariance}) for
the \textit{triholomorphic moment maps} reduces to
\begin{eqnarray}
-\varepsilon^{xyz}\,\mathcal{P}^y_\Lambda\,\mathcal{P}^z_\Sigma &=&
f^\Gamma{}_{\Lambda\Sigma}\,\mathcal{P}^x_\Gamma\,. \label{cocycond}
\end{eqnarray}
In the case $\mathcal{G}=\mathrm{SO(3)}$, $f^x{}_{yz}=e\,
\varepsilon_{xyz}$ ($e$ being the coupling constant) and the above
condition is satisfied by setting
\begin{equation}
 \mathcal{P}^x_{\Sigma}=  \left \{\begin{array}{lcl}
  - e\,\delta^x_{y} & \mbox{for} & \Sigma= 3+y \\
   0 & \mbox{for} &  \Sigma= \mbox{otherwise.} \
 \end{array} \right.
\label{fitermi}
\end{equation}
For each abelian $\mathrm{U(1)}$ generator $t_{\Lambda_\odot}$ included
in the gauge algebra, (\ref{cocycond}) can instead be satisfied by
setting:
\begin{equation}
  \mathcal{P}^x_{\Sigma}=  \left \{\begin{array}{lcl}
   e\,\delta^x_{3} & \mbox{for} & \Sigma= \Sigma_\odot \\
   0 & \mbox{for} &  \Sigma= \mbox{otherwise.} \
 \end{array} \right.
\label{abefitermi}
\end{equation}
In the context of conformal tensor calculus, the Fayet--Iliopoulos terms
represent the transformation of the compensating hypermultiplet under a
$\U(1)$ or $\SO(3)$ gauge group~\cite{deWit:1984pk}.

\subsection{General form of the scalar potential.}

Having introduced the above ingredients, we can apply the general formula
for the scalar potential of an $\mathcal{N}=2$ supergravity that was
derived in \cite{Andrianopoli:1997cm}. In order to write down such a
formula, we still need to recall one more definition. Given the
covariantly holomorphic section $V$ of special geometry (rotated in the
style of de Roo--Wagemans or not), we name
\begin{equation}
  f^\Lambda_\alpha  \, \equiv \, \left( \partial _\alpha  +\ft 1 2 \partial _\alpha
  \mathcal{K}\right)  L^\Lambda
\label{flamdefi}
\end{equation}
the K{\"a}hler-covariant derivatives of the upper electric part and we
introduce the positive-definite matrix
\begin{eqnarray}
{U}^{\Lambda \Sigma} \equiv g^{\alpha \bar \beta} f^\Lambda_\alpha
f^\Sigma_{\bar \beta} &=&  -\ft{1}{2} \left( \mbox{Im}\mathcal{N}\right)
^{-1 \vert \Lambda\Sigma} -
   \overline{L}^\Lambda \, {L}^\Sigma\,,\label{formulaU}
\end{eqnarray}
where $\mathcal{N}_{\Lambda\Sigma}$ is the kinetic matrix of the vector
fields.

With the normalization of the Lagrangian as
\begin{equation}
  e^{-1}{\cal L}= g_{\alpha \bar \beta }\nabla _\mu z^\alpha \nabla ^\mu z^{\bar \beta
  }+ h_{uv}\nabla _\mu q^u \nabla ^\mu q^v -V+\ldots \,,
 \label{normalL}
\end{equation}
the mass matrices for the special K{\"a}hler and quaternionic-K{\"a}hler parts
are
\begin{equation}
  (m^2)_\alpha {}^\beta = g^{\beta \bar \beta }\partial _\alpha \partial _{\bar \beta
  }V\,, \qquad  (m^2)_u{}^v= \ft12 h^{vw}\partial _u\partial _w V\,.
 \label{normalm}
\end{equation}
The scalar potential takes the form \cite{Andrianopoli:1997cm}
\begin{eqnarray}
\mathcal{V}&=&\left (g_{\alpha \bar \beta}k^\alpha_\Lambda k^{\bar
\beta}_\Sigma +4 h_{uv} k^u_\Lambda k^v_\Sigma \right ) \bar L^\Lambda
L^\Sigma+\left ( U^{\Lambda\Sigma}-3 \bar L^\Lambda L^\Sigma \right )
\mathcal{P}^x_\Lambda \mathcal{P}^x_\Sigma \nonumber\\
& = & \mathcal{V}_1 + \mathcal{V}_2 + \mathcal{V}_3\,, \label{ssette}
\end{eqnarray}
and is the sum of three distinct contributions:
\begin{eqnarray}
\mathcal{V}_1  & = & g_{\alpha \bar \beta}k^\alpha_\Lambda k^{\bar \beta }_\Sigma \,  \bar L^\Lambda L^\Sigma
=\rmi \left( M_\Delta f_{\Lambda \Gamma }{}^\Delta \bar L^\Lambda\right)
 \left( L^\Sigma f_{\Sigma \Pi }{}^\Gamma \bar L^\Pi \right)  +\hc\,,\nonumber\\
\mathcal{V}_2 & = & 4 h_{uv} k^u_\Lambda k^v_\Sigma \, \bar L^\Lambda
L^\Sigma\,,
\nonumber\\
\mathcal{V}_3 & = & \left ( U^{\Lambda\Sigma}-3 \bar L^\Lambda L^\Sigma
\right ) \, \mathcal{P}^x_\Lambda \mathcal{P}^x_\Sigma\,. \label{treadde}
\end{eqnarray}
By their definition, the contributions $\mathcal{V}_{1,2}$ are positive
definite and the only term that might involve negative contributions is
$\mathcal{V}_3$. This can be understood from the fact that
$\mathcal{V}_{1}$ and the first term constitute the square of the
supersymmetry transformation of the gauginos (split in the $\SU(2)$
triplet and $\SU(2)$ singlet part), $\mathcal{V}_2$ is the square of the
supersymmetry of the hyperinos, and the last term of $\mathcal{V}_3$ is
the square of the gravitino supersymmetry. It is well known that the
potential can be split in such a way, and that then only the gravitino
contribution is negative definite. The term $\mathcal{V}_1$ differs from
zero only if the gauge group is non-Abelian, as only then scalars of the
vector multiplets transform under $\mathcal{G}_{\rm gauge}$. Abelian
factors are characterized by vanishing Killing vectors and do not
contribute to $\mathcal{V}_1$. In the term $\mathcal{V}_2$, we have used
$k^u_\Lambda$ that are the Killing vector fields describing the action of
the gauge group on the quaternionic scalars pertaining to
hypermultiplets. Hence $\mathcal{V}_2$ is identically zero in the absence
of hypermultiplets. Finally, the crucial term $\mathcal{V}_3$ contains
the matrix $\left ( U^{\Lambda\Sigma}-3 \bar L^\Lambda L^\Sigma \right )$
which is made of the complex scalars sitting in the vector multiplets and
$\mathcal{P}^x_\Lambda$ that are the triholomorphic moment maps for the
action of the gauge group on the quaternionic scalars. So $\mathcal{V}_3$
describes contact interactions between the vector and the hyper scalars.
Due to Fayet--Iliopoulos terms, the term $\mathcal{V}_3$ can be non-zero
also in the absence of hypers and then takes contributions only if we
have $\mathrm{SO(3)}$ or $\mathrm{U(1)}$ factors.

\subsection{Abelian gauging in special geometry lead to tachyonic vacua}
\label{ss:AbelianDS}

To illustrate the difficulties for having stable de Sitter vacua, we give
the eigenvectors that were found in~\cite{deWit:1984pk,Cremmer:1985hj},
having negative masses related to the value~(\ref{eigenvaluem2}) as
recently found in higher $\mathcal{N}$ supergravities
in~\cite{Kallosh:2001gr}.

For Abelian gaugings involving only vector multiplets, the relevant
quantities are the contributions to the Fayet--Iliopoulos terms for each
generator, as in~(\ref{abefitermi}), i.e.,
\begin{equation}
  {\cal P}^x_\Lambda =g_\Lambda \delta ^{x3}\,.
 \label{FIAbelian}
\end{equation}
The potential again gets only contributions from $\mathcal{V}_3$
in~(\ref{treadde}), and can now be rewritten as
\begin{equation}
  {\cal V}=g^{\alpha \bar \alpha }{\cal D}_\alpha W\, {\cal D}_{\bar \alpha }\bar
  W -3 |W|^2\,,\qquad W\equiv g_\Lambda L^\Lambda\,.
 \label{VAbelian}
\end{equation}
The basic relations of special geometry imply
\begin{equation}
{\cal D}_\alpha {\cal D}_{\bar \beta } \LX^\ILambda = g_{\alpha\bar \beta
}\LX^\ILambda\,,\qquad
  {\cal D}_\alpha {\cal D}_\beta \LX^\ILambda = C_{\alpha \beta \gamma }
{\cal D}^\gamma
  \bar \LX^\ILambda\,,
 \label{relationscov}
\end{equation}
where $C_{\alpha \beta \gamma }$ is the covariantly holomorphic symmetric
3-index tensor appearing in the fundamental curvature relation:
\begin{equation}
  R_{\bar \alpha \beta \bar \gamma \delta }=2g_{\bar \gamma \beta }g_{\delta \bar \alpha
  }+2g_{\bar \gamma \delta }g_{\beta \bar \alpha }-2C_{\bar \alpha \bar \gamma \bar \epsilon }
  C_{\beta \delta \epsilon }g^{\bar \epsilon \epsilon }\,.
 \label{RiemannSG}
\end{equation}
Applying~(\ref{relationscov}) leads, as condition for extrema of the
potential, to
\begin{equation}
  \partial _\alpha V=C_{\alpha \beta \gamma }
{\cal D}^\beta \bar W\,{\cal D}^\gamma  \bar W
 -2\bar W {\cal D}_\alpha W=0\,.
 \label{dVSKa}
\end{equation}
There are first the solutions with ${\cal D}_\alpha W=0$, which lead to
the anti-de Sitter vacua. In other cases, we can use
\begin{equation}
2\bar W  |{\cal D}_\alpha W|^2= C_{\alpha \beta \gamma }{\cal D}^\alpha
\bar W\,{\cal D}^\beta \bar W\,{\cal D}^\gamma \bar W
 \label{valueDW2}
\end{equation}
as expression for $W$ at the extremum. This allows us to write the vacuum
value of the potential as
\begin{equation}
 V=|{\cal D}_\alpha W|^2-\frac{3}{4}\frac{\left|C_{\alpha \beta \gamma }{\cal D}^\alpha
\bar W\,{\cal D}^\beta \bar W\,{\cal D}^\gamma \bar W\right|^2}{|{\cal
D}_\alpha W|^4}\,.
 \label{valueVinC}
\end{equation}
The second derivative of the potential is for the
holomorphic--holomorphic and the holomorphic--antiholomorphic parts
\begin{eqnarray}
{\cal D}_\alpha \partial _\beta V&=& \left({\cal D}_\alpha C_{\beta
\gamma \delta }\right) {\cal D}^\gamma \bar W\,{\cal D}^\delta \bar W\,,\nonumber\\
 \partial ^\beta\partial _\alpha V & = & 2C_{\alpha \gamma \delta }\bar C
^{\beta \gamma \epsilon }{\cal D}_\epsilon W\,{\cal D}^\delta \bar
W-2{\cal D}^\beta \bar W
 \,{\cal D}_\alpha W-2\delta_\alpha ^ \beta |W|^2 \,.
 \label{VppC}
\end{eqnarray}
We get then for the trace with~(\ref{valueDW2})\footnote{We use here $n$
for the number of vector multiplets. The remainder of the paper uses
specifically the manifolds~(\ref{spiffero}), which defines the meaning of
$n$, and where the number of vector multiplets is, therefore, $n+1$.}
\begin{eqnarray}
 g^{\alpha \bar \beta }\partial _{\bar \beta }\partial _\alpha V  &=&-2|{\cal D}
 W|^2-\frac{n}{2}\frac{\left|C_{\alpha \beta \gamma }{\cal D}^\alpha
\bar W\,{\cal D}^\beta \bar W\,{\cal D}^\gamma \bar W\right|^2}{|{\cal
D}_\alpha W|^4} + 2C_{\alpha \gamma \delta }\bar C^{\alpha \gamma
\epsilon }{\cal D}_\epsilon W\,{\cal D}^\delta \bar W \nonumber\\
&=& -2\left(\frac14 V_0\right) -\frac{(3+n)}{2}\frac{\left|C_{\alpha
\beta \gamma }{\cal D}^\alpha \bar W\,{\cal D}^\beta \bar W\,{\cal
D}^\gamma \bar W\right|^2}{|{\cal D}_\alpha W|^4} + 2C_{\alpha \gamma
\delta }\bar C^{\alpha \gamma \epsilon }{\cal D}_\epsilon W\,{\cal
D}^\delta \bar W \,.\label{resultwithC}
\end{eqnarray}
Note that for $n=1$ (and thus all indices are the same), the last two
terms cancel. This implies that the trace of the eigenvalues of the mass
matrix is $-2$ in that case. It is a complex scalar; thus, there are two
eigenvalues. The separate eigenvalues are dependent on the actual value
of $C$. This is actually (6.22) of~\cite{deWit:1984pk}.

The result of~\cite{Cremmer:1985hj} that there is always an eigenvalue
$-2V$ in the holomorphic--anti\-holo\-mor\-phic derivative, is derived as
follows. One multiplies the second equation of~(\ref{VppC}) with ${\cal
D}_\beta W$ and uses~(\ref{dVSKa}) and its complex conjugate. This leads
to
\begin{equation}
  \partial ^\beta \partial _\alpha V\,{\cal D}_\beta W= -2 V\,{\cal D}_\alpha
  W\,.
 \label{eigenvalue-2}
\end{equation}
This shows that for Abelian gaugings with only vector multiplets, stable
de Sitter vacua cannot exist, because we always have a complex tachyon
with characteristic negative mass as in~(\ref{eigenvalue-2}). Henceforth,
the possibility of finding stable de Sitter vacua relies on the
contribution $\mathcal{V}_1$ coming from non-Abelian gaugings. In the
following section, we prove that this can be sufficient to produce the
desired result.
\section{The three models}\label{ss:3models}
The three models that we are going to present are
\begin{itemize}
  \item a model with 3 vector multiplets, in the manifold $\mathcal{ST}[2,2]$,
  which, together with the graviphoton, are gauging
  $\SO(2,1)\times \U(1)$, with a Fayet--Iliopoulos term for the $\U(1)$.
  \item a model with 5 vector multiplets, in the manifold $\mathcal{ST}[2,4]$,
  which, together with the graviphoton, are gauging
  $\SO(2,1)\times \SO(3)$, with a Fayet--Iliopoulos term for the
  $\SO(3)$; and
  \item the last model extended with 2 hypermultiplets with 8 real
  scalars in the coset $\frac{\SO(4,2)}{\SO(4)\times \SO(2)}$.
\end{itemize}
In the choice of the hypermultiplet sector, we made use of the fact that
we can use the coset $\frac{\SO(4,2)}{\SO(4)\times \SO(2)}$ as well as a
factor in the special K{\"a}hler manifold $\mathcal{ST}[2,4]$ as for a
quaternionic-K{\"a}hler manifold. Moreover, as we will discuss in the last
section, such a choice makes a first step towards a generalization to
$\mathcal{N}=4$ supergravity and string theory. We will now discuss the
three models consecutively.

\subsection{$\SO(2,1)\times \U(1)$ gauging}

In this case, the Cartan--Killing metric on the group manifold,
$\eta_{(2,2)}={\diag}(+,+,-,-)$, naturally splits into the
Cartan--Killing metric of the first non-Abelian non-compact factor,
namely $\eta_{(2,1)}={\diag}(+,+,-)$ of $\SO(2,1)$ plus
$\eta_{(1)}={\diag}(-)$. In general, we denote by $e_0$ and $e_1$ the
coupling constants of the non-compact and compact factors, respectively.
Since $\U(1)$ is Abelian, $e_1$, rather than a coupling constant, is
actually just the value of the Fayet--Iliopoulos parameter. The first
three of the four Killing vectors $k_\Lambda$ corresponding to the
generators of the gauge group $\SO(2,1)$ are given in~(\ref{so21kil}),
while the fourth one is zero. The scalar potential has the form
\begin{eqnarray}
\mathcal{V}(S,\,\bar{S},y,\bar{y})&=&g_{\alpha\bar  \beta
}\,k_x^\alpha\,k_y^{\bar
\beta }\,\bar{L}^x\,L^y +\mathcal{V}_3\,,\label{pot2ter}\\
\mathcal{V}_3&=&(U^{\Lambda\Sigma}-3 \bar L^\Lambda L^\Sigma )
\mathcal{P}_\Lambda^x \mathcal{P}_\Sigma^x =e_1^2\, \left(U^{{4}\,{4}}-3
\bar L^{4} L^{4}\right).\label{u1fiter}
\end{eqnarray}
In the first term of (\ref{pot2ter}), only the terms with $x=1,2,3$
contribute, due to the vanishing of the $\U(1)$ Killing vector. In the
Fayet--Iliopoulos term of (\ref{u1fiter}) instead, the only contribution
comes from $\Lambda=\Sigma=4$. This is so, because, for an Abelian group
the constant moment map is
\begin{equation}
  \mathcal{P}_\Sigma^x =  e_1\,\delta ^{x3} \, \delta _{\Lambda 4}\,.
\label{Pabelian}
\end{equation}
Without de Roo--Wagemans rotation ($a_1 = 0$), the matrix appearing in
the $\mathcal{V}_3$ would take the simple form (see (9.58) of
~\cite{Andrianopoli:1997cm})
\begin{eqnarray}
U^{\Lambda\Sigma}-3 \bar L^\Lambda L^\Sigma &=& \frac{1}{2\Im
S}\,\eta_{(2,2)}^{\Lambda\Sigma}\,,
\end{eqnarray}
and therefore the $\mathcal{V}_3$ term would be
\begin{eqnarray}
\mathcal{V}_3&=&e_1^2\, \left(U^{{4}\,{4}}-3 \bar L^{4}
L^{4}\right)=\frac{1}{2\Im S}e_1^2\, \eta_{(2,2)}^{{4}\,{4}}=-
\,\frac{1}{2\Im S}e_1^2>0\,,
\end{eqnarray}
where we have used the property ${\rm Im}(S) < 0$ required by ${\rm
Im}(\mathcal{N}) < 0$. This potential has obviously no extremum, and here
the de Roo--Wagemans rotation~(\ref{dRWagangles}) (with
$\theta=\theta_1$) comes to rescue. Indeed, the effect of such a rotation
amounts to a modular transformation of $S$, so that
\begin{eqnarray}
\mathcal{V}_3&=&-e_1^2\frac{1}{2\Im S}\,|\cos{(\theta
)}-S\,\sin{(\theta)}|^2>0 \,.\label{sintre}
\end{eqnarray}
\par
Let us now study the critical point of this potential. Explicitly, we
obtain the following form:
\begin{equation}
\mathcal{V}_{\SO(2,1)\times \U(1)}=\mathcal{V}_3+\mathcal{V}_1=
-\frac{1}{2\Im S} \,\left(e_1{}^2 |\cos \theta -S\,\sin \theta|^2 +
      e_0{}^2\,\frac{{P_2^+}(y)}{{P_2^-}(y)}
      \right),
\label{Potentabel}
\end{equation}
where $P_2^\pm  (y) $ are polynomial functions in the Calabi--Vesentini
variables of the holomorphic degree specified by their index\footnote{We
write from now on explicit components of the $y^a$ variables with lower
indices to distinguish them from squares of sums, \ldots .} (here only 2,
but will be higher in the next example)
\begin{equation}
P_2^\pm (y)  =  1 - 2\,y_0\,\overline{y}_0\pm
  2\,y_1\,{\overline{y}_1} +y^2\bar y^2\,.
\label{polinabel}
\end{equation}
The last term in~(\ref{Potentabel}) is another positive definite term
that originates from the non-Abelian non-compact $\SO(2,1)$ gauging.
Indeed, this term is just the norm of the Killing vectors.

\begin{figure}[ht]
\unitlength=1mm
\begin{center}
\begin{picture}(80,80)(0,0)
\put(10,0){\leavevmode \epsfxsize=6cm
 \epsfbox{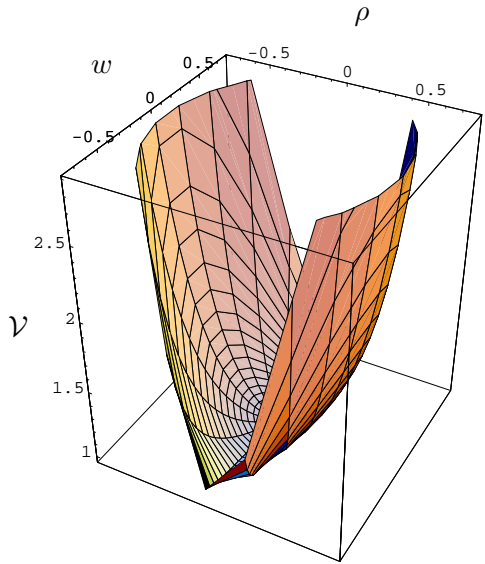}}
\put(53,72){$\rho $} \put(18,65){$w$} \put(7,30){$\mathcal{V}$}
\end{picture}
\caption{\it Potential~(\ref{VatS0}) in units of $|e_0e_1\sin\theta |$
for $\delta =\pi /10$. For different values of $\delta $, the picture
maintains the same shape.\label{fig:potenfig}}
\end{center}
\end{figure}

We now look for an extremum. Equating to zero the $S$ derivative, we
obtain, consistent with the positivity of the kinetic term of the vector
fields,
\begin{equation}
S=S_0= \cot \theta -\rmi \frac{e_0}{e_1}\frac{1}{\sin\theta
}\sqrt{\frac{P_2^+(y)}{P_2^-(y)}}\,, \label{Svalue}
\end{equation}
where we assumed that $e_0 e_1 >0$. Inserting this extremum value of $S$
in the potential~(\ref{Potentabel}), we obtain
\begin{equation}
\left.\mathcal{V}_{\SO(2,1)\times \U(1)}\right|_{S=S_0}=
e_0e_1\sin\theta\sqrt{\frac{P_2^+(y)}{P_2^-(y)}}=(-\Im
S_0)\,e_1^2\,\sin^2\theta\,.
 \label{VatS0}
\end{equation}
At the level of the above equation, where the extremum is only considered
in the $S$-direction, we already reach the very relevant conclusion that
the potential is strictly positive in the positivity domain of the
Lagrangian ($\Im S<0$) and might vanish only at the boundary of moduli
space. Therefore, we can have only de Sitter and no Minkowski vacua.
Setting $y_0=w\exp \rmi\beta $ and $y_1=\rho \exp\rmi(\beta +\delta ) $,
the polynomials are
\begin{equation}
  P_2^\pm = -2w^2+w^4+\left( \pm 1+\rho ^2\right) ^2+2w^2\rho ^2\cos 2\delta
  \,.
 \label{P2inwrhodelta}
\end{equation}
The resulting potential is illustrated in figure~\ref{fig:potenfig}. It
displays a valley along the direction of $w$, for $\rho =0$. Note that
for $\rho =0$, we have $L^\Lambda P^x_\Lambda =0$, which means that the
gravitino shift (transformation of the gravitino under supersymmetry due
to the scalars) is zero, as we will further discuss in the next model.
This will, of course, also lead to a zero mode in the mass matrix.

The critical values of the $y$ fields are
\begin{eqnarray}
y_0  =  \mbox{arbitrary} \,,\qquad  y_1  =  0 \,,\qquad \mbox{i.e. }\
\rho =0\,. \label{ynoteval}
\end{eqnarray}
Therefore, the extremum value of the potential is at
\begin{equation}
\mathcal{V}=\mathcal{V}_0 =  e_0\,e_1\,\sin\theta  > 0\,. \label{V_0}
\end{equation}
The matrix of second derivatives of the potential at the extremum,
normalized using the inverse metric
\begin{equation}
  g^{S\bar S}=4(\Im S)^2\,,\qquad \left.g^{a\bar b}\right|_{0}=\ft12\delta ^{ab}
  (1-w^2)^2\,,
 \label{ginverse}
\end{equation}
is the mass matrix
\begin{equation}
\left.\frac{\partial _\alpha \partial^ \beta
\mathcal{V}}{\mathcal{V}}\right|_{0}= \left(  \matrix{ 2 & 0 & 0 \cr
        0 & 0 &  0 \cr
        0 & 0 &  1 \cr  }\right)  \,,  \qquad
          \left.\partial _\alpha \partial _{ \beta }\mathcal{V}\right|_{0}=
          0\,.
\label{massamtrinorm}
\end{equation}
This displays two (complex) positive and one (complex) null eigenvalue.
The zero modes are the two Goldstone bosons of the non-compact
translations that become massive and are also the moduli of the flat
direction displayed by the potential. We will explicitly show this in the
next model, which has the same essential properties as the present toy
model.

Thus, we find a stable de Sitter vacuum, with characteristic squared mass
values for the scalars.


\subsection{$\SO(2,1)\times \SO(3)$ gauging}
\label{so21so3}
 Let us now work out in detail the case of $\mathcal{N}=2$
supergravity coupled to 5 vector multiplets whose scalar components span
the manifold $\mathcal{ST}\left[2,4\right]$ choosing as gauge group
$\mathcal{G}_{\rm gauge}=\mathrm{SO(2,1)}\times \mathrm{SO(3)}$ and
including a FI term associated with the second factor. In this case, the
metric on the vector field representation
$\eta_{(2,4)}=\diag(+,+,-,-,-,-)$ naturally splits into the
Cartan--Killing metrics of the two factor groups, namely
$\eta_{(2,1)}=\diag(+,+,-)$ and $\eta_{(3)}=\diag(-,-,-)$. We denote by
$e_0$ and $e_1$ the coupling constants of the non-compact and compact
factor, respectively. The Killing vectors $k_\Lambda$ corresponding to
the generators of the gauge group are therefore given by (\ref{so21kil})
and (\ref{so3killi}).
\par
Due to the absence of hypermultiplets, $\mathcal{V}_2$ is still not
contributing to the potential. The $\mathcal{V}_3$ term, instead, is
\begin{eqnarray}
\mathcal{V}_3&=&(U^{\Lambda\Sigma}-3 \bar L^\Lambda L^\Sigma )
\mathcal{P}_\Lambda^x \mathcal{P}_\Sigma^x =e_1^2\,
\sum_{x=1}^3\,\left(U^{(x+3)(x+3)}-3 \bar L^{x+3} L^{x+3}\right)\nonumber\\
&=&\frac1{2\Im S }\,e_1^2\,|\cos\theta -S\sin\theta |^2\,
\sum_{x=1}^3\,\eta_{(2,4)}^{(x+3)(x+3)}=-\frac3{2\Im S
}\,e_1^2\,|\cos\theta -S\sin\theta |^2>0\,.\label{thatsit}
\end{eqnarray}
As one notices, the only difference between the Abelian and non-Abelian
case in the Fayet--Iliopoulos term is a factor 3. Each generator of
$\SO(3)$ gives the same contribution as the $\U(1)$ generator in the
Abelian case.

Also in this case, the potential is positive definite since both
$\mathcal{V}_1$ and $\mathcal{V}_3$ are positive definite. Their sum
gives
\begin{equation}
  \mathcal{V} = -\frac{1}{2\Im S}\left[ e_1{}^2 \, |\cos \theta  -
S\,\sin\theta |^2\,\,\frac{P_4^{(1)}}{P_2{}^2} +
  e_0{}^2\,\frac{P_4^{(0)}}{P_2{}^2}\right].
\label{pot}
\end{equation}
where $ P_\ell (y,\,\bar{y})$ ($\ell=2,\,4$) are polynomials of
holomorphic degree $\ell$ in $y$, whose important properties are
\begin{eqnarray}
  &&P_2=1-2y\bar y+y^2\bar y^2\,, \nonumber\\
  &&\left.\partial_{y^a}\,P_\ell\right|_{y= 0}=0\,,\qquad
  \left.P_4^{(0)}\right|_{y= 0}=1\,,\qquad \left.P_4^{(1)}\right|_{y= 0}= 3\,. \label{propP}
\end{eqnarray}
Setting to zero the $S$-derivative of the scalar potential in
(\ref{pot}), we obtain the critical value of $S$ in terms of the $y$
fields
\begin{equation}
  S=S_0(y)=\cot \theta -\rmi\,
\left|\frac{e_0}{\,e_1}\frac{1}{\sin\theta
}\right|\sqrt{\frac{P_4^{(0)}(y)}{P_4^{(1)}(y)}}\,.
 \label{S0model2}
\end{equation}
Inserting this in the potential, reduces it to
\begin{equation}
  \left.\mathcal{V}\right|_{S=S_0}=|e_0\,e_1\,\sin\theta |\sqrt{\frac{P_4^{(0)}(y)\,P_4^{(1)}(y)}{P_2^4(y)}}
\,.
 \label{VatS0model2}
\end{equation}
With the properties~(\ref{propP}), we conclude that the potential reaches
an extremum at
\begin{equation}
z=\phi^{(0)}=\{S_0(0),y_0=0\}\,,\qquad S_0(0)= \cot \theta -\rmi\,
\left|\frac{1}{\sqrt{3}}\,\frac{e_0}{e_1}\,\frac{1}{\sin\theta
}\right|\,.
\end{equation}
At this extremum, the potential has the  value
\begin{equation}
\mathcal{V}_0=\mathcal{V}(\phi^{(0)})= \sqrt{3}\,|e_0\,e_1\,\sin\theta |>
0\,. \label{V0inmodel2}
\end{equation}
Hence this extremum defines again a de Sitter space.
\par
\paragraph{More detailed analysis of the potential.}
The de Sitter vacuum that we displayed is the only possibility in the
positivity domain of the Lagrangian. Just as in the first example, the
extremum we have found is actually a point on a full line of extrema.
This can be seen in 2 ways. Either by calculating the mass matrix and
showing that it involves a zero mode, or, alternatively, the presence of
a flat direction can be appreciated through a more detailed analysis of
the potential. This involves a closer look at the structure of the
polynomial functions $P_4^{(0,1)}(y,{\bar y})$ and $P_2(y,{\bar y})$
appearing in the final form (\ref{pot}) of the potential. This is what we
do in this paragraph.

Since the potential is $\mathrm{SO(3)}$ invariant the best choice of
variables are $\mathrm{SO(3)}$ invariants. We set
\begin{eqnarray}
y_0  =  w \,\rme^ {{\rm i} \,\beta }\,, \qquad \vec{\mathbf{y}}  =  \vec
{\mathbf{v}}_1 +{\rm i} \, \vec {\mathbf{v}}_2 \,,\label{v1v2vectors}
\end{eqnarray}
and expect that the polynomial functions entering the potential should
depend only on
\begin{equation}
|\mathbf{v}_1|^2  \equiv  \rho^2 \, \cos \phi \,,\qquad |\mathbf{v}_2|^2
 \equiv  \rho^2 \, \sin \phi  \,,\qquad  \vec {\mathbf{v}}_1 \, \cdot \,
\vec {\mathbf{v}}_2  \equiv  \rho^2 \sin \phi \cos \phi  \, \cos\theta
\,. \label{luccolo}
\end{equation}
Indeed this is what happens, and by means of a Mathematica program, one
finds an explicit form of the $P_\ell $ polynomials, which is too lengthy
to display here in full generality. Inserting the value~(\ref{S0model2})
of the $S$ field into the derivatives of the potential $\mathcal{V}$ with
respect to $y^a$, the following equation should hold:
\begin{equation}
  0= F_a \, \equiv \, P_2 \, P_4^{(1)} \, \partial _a P_4^{(0)} +
 P_4^{(0)} \, \left( -4 \,P_4^{(1)} \,
 \partial _a P_2 + P_2 \, \partial _a P_4^{(1)}\right).
\label{polycond}
\end{equation}
An $\SO(3)$-invariant vacuum should occur at $\vec y=0$ (i.e. $\rho =0$).
For this choice, the polynomials simplify dramatically and we obtain
\begin{equation}
  P_0|_{\rho=0} =(-1+w^2)^2\,, \qquad P_4^{(0)}|_{\rho=0} =
  (-1+w^2)^4 \,,\qquad  P_4^{(1)}|_{\rho=0} = 3 \, (-1+w^2)^4\,.
\label{zerorhovalues}
\end{equation}
Inserting this in~(\ref{VatS0model2}), the potential becomes the constant
$\mathcal{V}_0$ of~(\ref{V0inmodel2}) independent of $w$ and $\beta$. In
other words, we have an extremum for arbitrary values of $y_0$ as
claimed.

\paragraph{The mass matrix.} The factors of 3 and $\sqrt{3}$ that in this
model are extra with respect to the first model, disappear in the final
mass matrix, and we find
\begin{equation}
\left.\frac{\partial _\alpha \partial^ \beta
\mathcal{V}}{\mathcal{V}}\right|_{0}= \left(  \matrix{ 2 & 0 & 0 \cr
        0 & 0 &  0 \cr
        0 & 0 &  \unity _3 \cr  }\right)  \,,  \qquad
          \left.\partial _\alpha \partial _{ \beta }\mathcal{V}\right|_{0}=
          0\,.
\label{massamtrinorm2}
\end{equation}
We thus find the same conclusions as in the previous model, and the shape
of the potential is also similar to the one in figure~\ref{fig:potenfig}.

\paragraph{Supersymmetry breaking.}
Let us now investigate the supersymmetry of the solution. The
supersymmetry variations of the gravitinos and gauginos are
\begin{equation}
\delta \psi_{i\mu}=\rmi S_{ij} \gamma_\mu \epsilon^j\,,\qquad
\delta\lambda^{i\alpha }=  W^{\alpha ij}\epsilon_j\,,
\label{susyFermTrans}
\end{equation}
where we have
\begin{eqnarray}
&&S_{ij}=-\ft12{\rm i} {\cal P}_{\Lambda ij}L^\Lambda\,,\qquad W^{\alpha
ij}=\varepsilon^{ij}\,k_{\Lambda}^\alpha   \bar L^\Lambda\,-\, {\rm
i}{\cal P}^{ij}_{\Lambda} g^{\alpha \bar \beta } {\bar f}_{\bar \beta
}^{\Lambda}\,,\nonumber\\ && {\cal P}_{\Lambda
ij}\equiv(\sigma_x)_i{}^k\varepsilon_{kj} {\cal P}^x_{\Lambda}\,, \qquad
{\cal P}^{ij}_\Lambda\equiv({\cal P}_{\Lambda ij})^*\,. \label{defSandW}
\end{eqnarray}
Since at the extremum ($\vec y=0$) we have $L^{x+3}=0$, we find that the
gravitino mass matrix $S_{ij}$ vanishes there. The gaugino shifts,
instead, are
\begin{eqnarray}
W^{S\,ij}&=&0\,,\qquad
W^{0\,ij}=-\rmi\frac{e_0}{2\,\sqrt{-\Im S_0(0)}}\,\varepsilon ^{ij}\,,\nonumber\\
W^{x\,ij}&=&-\rmi\frac{e_1}{2\,\sqrt{-\Im S_0(0)}}\,\left(\cos \theta  -
{{\bar{S}}}\,\sin\theta \right)\,(\sigma_x)^{ij}\,, \label{gauginoshift}
\end{eqnarray}
where the indices $S$, 0 and $x$ enumerate the basis of scalar fields
$z^\alpha=\{S,y_0, y_x\}$. Since the gaugino shift
matrices~(\ref{gauginoshift}) do not have any common zero eigenvalue,
supersymmetry is broken to $\mathcal{N}=0$.

\paragraph{Vector fields and BEH effect.}

The imaginary part of the {\it period matrix} has the form
\begin{equation}
\mathcal{N}-\bar{\mathcal{N}}=2\rmi\Im S\left(\matrix{M_3 & 0\cr 0&
\frac{1}{|\cos\theta - S\,\sin\theta |^2}\,\unity _3}\right)
\end{equation}
where the $3\times 3$ matrix $M_3$ has been calculated by Mathematica and
has all non-zero entries for $y_0\neq 0$. Although we can calculate the
effective vector field Lagrangian in a generic extremum $y_0\neq 0$ by
diagonalizing $M_3$, for the sake of simplicity we just consider $y_0=0$
where $M_3=\unity _3$. Then we can show that the two gauge vectors
associated with the non-compact generators $T_{\Lambda=1,2}$ of
$\SO(2,1)$ acquire dynamical masses consistently with the Goldstone
theorem. Indeed, these two generators are broken by our vacuum since by
looking at the Killing vectors~(\ref{so21kil}) we see that $y_0$ is not
invariant under these two generators, and we rather have
\begin{equation}
\delta_{\Lambda=1} \,y_0= -\ft12\rmi e_0\,,\qquad \delta_{\Lambda=2}
\,y_0=\ft12 e_0\,.
\end{equation}
Hence, the residual symmetry of all the de Sitter extrema that we have
constructed is $\SO(2)\times \SO(3)$, where $\SO(2)$ is the compact part
of $\SO(2,1)$. The masses of the vector fields arise from the kinetic
term for the scalar fields $\nabla_\mu z^\alpha \,\nabla^\mu
\bar{z}^{\bar \beta }\,g_{\alpha \bar \beta }$ where $\nabla_\mu z^\alpha
=\partial_\mu z^\alpha  +A^\Lambda_\mu\, k^\alpha _\Lambda$. The mass
term at the extremum $\phi^{(0)}$ is
\begin{eqnarray}
A^\Lambda_\mu\, A^{\Sigma\,\mu}\,\left(k^\alpha _\Lambda\,k^{\bar \beta
}_\Sigma\,g_{\alpha \bar \beta}\right)_{z=\phi^{(0)}}&
=&\ft12e_0{}^2\,\left(A^1_\mu\, A^{1|\,\mu}+A^2_\mu\, A^{2|\,\mu}\right),
\end{eqnarray}
where we used the extremum value $g_{a\bar b}=2\delta _{ab}$. Hence, the
effective Lagrangian for the massive vector fields $A^{0,1}_\mu$ reads
\begin{eqnarray}
&&\Im S_0(0)\,\left(F^1_{\mu\nu}F^{1| \mu\nu}+F^2_{\mu\nu}F^{2|
\mu\nu}\right)+\ft12e_0{}^2\, \left(A^1_\mu\, A^{1|\,\mu}+A^2_\mu\,
A^{2|\,\mu}\right) \nonumber\\
&&= -\ft{1}{4}\, \left(\tilde{F}^1_{\mu\nu}\tilde{F}^{1|
\mu\nu}+\tilde{F}^2_{\mu\nu}\tilde{F}^{2| \mu\nu}\right)+
\ft12\mu^2\,\left(\tilde{A}^1_\mu\, \tilde{A}^{1|\,\mu}+\tilde{A}^2_\mu\,
\tilde{A}^{2|\,\mu}\right),
\nonumber\\
&&\mu^2\equiv -\frac{e_0^2}{4\,\Im S_0(0)
}=\frac{\sqrt{3}\,|\sin\theta\,e_0\,e_1|}{4}\,,
\end{eqnarray}
where we have redefined $\tilde{A}_\mu=2\,\sqrt{-\Im S_0(0)}\,A_\mu$. The
kinetic term of the Goldstone boson $y_0$ can be absorbed by a gauge
transformation on the broken gauge vector fields
\begin{equation}
A^{1}_\mu \rightarrow  A^{1}_\mu+\frac{2}{e_0}\,\partial_\mu\,\Im y_0\,,
\qquad A^{2}_\mu \rightarrow  A^{2}_\mu-\frac{2}{e_0}\,\partial_\mu\,\Re
y_0\,.
\end{equation}
In a vacuum where $y_0\neq 0$, the three vectors $A^{\Lambda=0,1,2}_\mu$
are mixed together. Two linear combinations become massive while there is
always one that remains massless.

\subsection{$\SO(2,1)\times \SO(3)$ gauging with hypers} \label{ss:withhyper}
Let us now generalize our previous results to an $\mathcal{N}=2$ model
with 5 vector multiplets and  4 hypermultiplets. We shall consider the
scalar fields in the vector- and hypermultiplets $S,\,y^a,\,q^u$ (with
$a=0,1,2,3$ and $u=1,\ldots ,8$) spanning the following product space:
\begin{eqnarray}
\mathcal{M}_{\mbox{\rm scal}}&=&
\left[\frac{\SU(1,1)}{\U(1)}\times\frac{\SO(2,4)}{\SO(2)\times
\SO(4)}\right]\times \left[\frac{\SO(4,2)}{\SO(4)\times \SO(2)}\right].
\label{MscalModel3}
\end{eqnarray}
The gauge group of our model is $G=\SO(2,1)\times \SO(3)$ which is
embedded in the $ \SO(2,4)$ subgroup of both the isometry group of the
special K{\"a}hler and of the quaternionic manifold. We can choose for each
factor whether this isometry is actually gauged (coupled to the vector
multiplets) or not. As we had chosen $e_0,\,e_1$ as the coupling
constants of $\SO(2,1)$ and $\SO(3)$, respectively, we now couple the
quaternionic scalars with factors $r_0\,e_0,\,r_1\,e_1$ where $r_0$ and
$r_1$ can be 0 or 1, indicating whether the corresponding isometry is
gauged or not. As we will see, we need the coupling to $\SO(3)$, as this
is replacing the FI term in the previous models, while for $r_0$ we can
consider both choices $r_0=0$ and $r_0=1$.
\par

It is convenient to choose the {\it solvable} parametrization of the
quaternionic manifold \cite{Andrianopoli:1997bq,Andrianopoli:1997zg}
\begin{eqnarray}
\frac{\SO(4,2)}{\SO(4)\times \SO(2)}&\equiv&\exp{\left({\rm Solv}\right)}\nonumber\\
{\rm Solv}&=&\sum_{u=1}^8\,q^u\,T_u\,=\,\nonumber\\&&a_1\,
E_{\epsilon_1-\epsilon_2}+a_2\,
E_{\epsilon_1+\epsilon_2}+\frac{a_3}{\sqrt{2}}\,\left(
E_{\epsilon_1+\epsilon_3}+ E_{\epsilon_1-\epsilon_3}\right)+ {\rm
i}\,\frac{a_4}{\sqrt{2}}\,\left( E_{\epsilon_1+\epsilon_3}-
E_{\epsilon_1-\epsilon_3}\right)+\nonumber\\&&
+\frac{b_1}{\sqrt{2}}\,\left( E_{\epsilon_2+\epsilon_3}+
E_{\epsilon_2-\epsilon_3}\right)+ {\rm i}\,\frac{b_2}{\sqrt{2}}\,\left(
E_{\epsilon_2+\epsilon_3}-
E_{\epsilon_2-\epsilon_3}\right)+h_1\,H_{\epsilon_1}+h_2\,H_{\epsilon_2}\,.
\label{solvable}
\end{eqnarray}
We started from the $6\times 6$ matrices in $\SO(4,2)$, using as Cartan
subalgebra the matrices
\begin{equation}
 H_{\epsilon_1}=M^{35}\,,\qquad H_{\epsilon_2}=M^{46}\,,\qquad
{\rm i}H_{\epsilon_3}=M^{12}\,,\qquad \mbox{where}\qquad
(M^{AB})^C{}_D\equiv 2\eta^{C[ A}\delta ^{B]}_D\,,
 \label{defCartan}
\end{equation}
with $A=1,\ldots ,6$ and the invariant metric of $\SO(4,2)$ is $\eta =
\eta^{(4,2)}=\diag\left(+,\,+,\,+,\,+,\,-,\,-\right)$. The $E$-matrices
in~(\ref{solvable}) are the $\SO(4,2)$ roots. The real form is such that
the $\epsilon _3$ direction is imaginary, while the $\epsilon _1$ and
$\epsilon _2$ are real. The generators that appear in~(\ref{solvable})
form the solvable algebra corresponding to this coset. These are
generators with non-negative $\epsilon _1$-weight. The scalars with
positive $\epsilon _1$-weight (the scalars $a_i$) are Peccei--Quinn
scalars, which will not enter the metric $h_{uv}$.

The coset representative is defined as follows:
\begin{eqnarray}
\mathbb{L}&=&\rme^{\sum_{i=1}^4\,a_i\,T_i}\cdot
\rme^{\sum_{k=1}^2\,b_k\,T_{k+4}}\cdot \rme^{\sum_{k=1}^2\,h_k\,
H_{\epsilon_k}}\,. \label{cosettusL}
\end{eqnarray}
It satisfies $\mathbb{L}^{-1}=\eta \mathbb{L}^T\eta $  and its explicit
expression is given in~\ref{bloccus}. This defines the vielbein 1-form,
which was the fundamental quantity in section~\ref{ss:QK}, as
\begin{equation}
  V^{mt}=(\mathbb{P}_4\, \mathbb{L}^{-1}\rmd \mathbb{L}\, \mathbb{P}_2)^{mt}\,,
 \label{defEfromcoset}
\end{equation}
where $\mathbb{P}_4$ and $\mathbb{P}_2$ are the projection matrices
splitting the range $A=1,\ldots ,6$ in $m=1,\ldots ,4$ and 5,6 being
re-labelled as $t=1,2$:
\begin{equation}
  \mathbb{P}_4=\pmatrix{\unity _4&0\cr 0&0}\,, \qquad \mathbb{P}_2=\pmatrix{0&0\cr 0&\unity
  _2}\,.
 \label{p4p2}
\end{equation}
The expression is given explicitly in~(\ref{Vmtexpl}), and the
corresponding metric, defined by~(\ref{quatmetric}), is given
in~(\ref{huv}). Observe that at the base point, where all the coordinates
$q^u=0$, the metric is just $h_{uv}=\ft12\delta _{uv}$. Also the $\SU(2)$
curvature is defined from the vielbein using~(\ref{SU2curv}), and the
connection 1-form in that equation can be written as\footnote{To check
these expressions and the ones below, it is useful to remark that
$\varepsilon ^{xyz}\mathbb{J}^x_{mn}\mathbb{J}^y_{pq}=2\left[ \delta
_{m[p}\mathbb{J}^z_{q]n}-(m\leftrightarrow n)\right]$. }
\begin{equation}
  \omega ^x=-\ft12 \mathbb{J}^x{}_m{}^n \left( \mathbb{P}_4 \,\mathbb{L}^{-1}\rmd
  \mathbb{L}\,\mathbb{P}_4\right) _n{}^m\,.
 \label{omegasolv}
\end{equation}

The isometries that are gauged by the vectors in the special K{\"a}hler
sector, i.e., $\SO(2,1)\times \SO(3)$, are acting on the {\bf 6} of
$\SO(4,2)$. Due to the structure of the $\eta ^{(4,2)}$, the $\SO(3)$
generators are chosen  to act on the first three components, while the
$\SO(2,1)$ act on the last three. See the explicit form of $t_\Lambda $
in~(\ref{gruppis}). The Killing vectors are then expressed as
\begin{equation}
  k_\Lambda ^u V_u^{mt}= \left( \mathbb{P}_4\,\mathbb{L}^{-1}t_\Lambda
  \mathbb{L}\, \mathbb{P}_2\right) ^{mt}\,.
 \label{trikillo}
\end{equation}
The corresponding tri-holomorphic momentum maps,
see~(\ref{introMomentMap}), have the following form:
\begin{eqnarray}
\mathcal{P}^x_\Lambda &=&\ft{1}{2}\,\mathbb{J}^x{}_m{}^n\left(
\mathbb{P}_4\,\mathbb{L}^{-1}t_\Lambda
  \mathbb{L}\, \mathbb{P}_4\right)_n{}^m\,. \label{Plam}
\end{eqnarray}
The explicit expressions of the Killing vectors and moment maps are quite
complicated in this parametrization, but are simpler in another
formalism.

Indeed, an alternative parametrization of the quaternionic coset manifold
consists in using the analogue $\tilde{y}^a$ of the Calabi--Vesentini
coordinates $y^a$ used to describe the $\SO(2,4)/\SO(2)\times \SO(4)$
factor in the special K{\"a}hler manifold. The relation between the
$\tilde{y}^a$ coordinates and the solvable ones
$a_i,\,b_1,\,b_2,\,h_1,\,h_2$ is highly nonlinear and is formally
discussed in appendix~(\ref{app:CV}).  Near the origin of the
quaternionic manifold, the relation between the corresponding
fluctuations can be linearized and will be used in the sequel for writing
the mass eigenstates in two relevant cases. Some quantities are most
conveniently expressed in one set of coordinates, the others in the other
set. For instance, in terms of $\tilde{y}^a$ the Killing vectors of the
gauge group have the same expression as those acting on the special
K{\"a}hler manifold provided one performs the obvious substitutions
$y^a\rightarrow \tilde{y}^a$, $e_0\rightarrow r_0\,e_0$ and
$e_1\rightarrow r_1\,e_1$. However, for the sake of constructing the
scalar potential, for which one needs an explicit expression of the coset
representative to work with, we find the solvable parametrization more
convenient.\par

Now we have the ingredients to compute the scalar potential $\mathcal{V}$
given by (\ref{ssette}). Although its expression is rather complicated,
one can immediately verify that
\begin{eqnarray}
&&\partial_y\mathcal{V}_{|y^a=q^u=0}=0\,,\qquad
\partial_q\mathcal{V}_{|y^a=q^u=0}=0\,,\nonumber\\
&&\mathcal{V}_{|y^a=q^u=0}= -\frac{1}{2 \Im S}\left[ 3 \,
r_1^2\,e_1^2\,|\cos \theta - S\,\sin\theta|^2+
e_0^2{\left(1+2\,r_0^2\right)}\right] \,.
\end{eqnarray}
{\it The expression of the potential at the origin of the quaternionic
manifold is the same as the one found in the previous models,
see~(\ref{Potentabel}) and~(\ref{pot}) at $y=0$, except that if the
$\SO(2,1)$ isometry is gauged, there is a rescaling in the $\SO(2,1)$
coupling constant: $e_0^2\,\rightarrow 3e_0^2$}.

A main ingredient is the triholomorphic moment map. This is related to
the complex structures acting in the upper $4\times 4$ part of the
$\SO(4,2)$ matrices, see~(\ref{Plam}). The generators of $\SO(2,1)$
within the isometry group of the quaternionic manifold are zero in this
upper $4\times 4$ block according to~(\ref{gruppis}). On the other hand,
those of $\SO(3)$ are chosen in this block. At the origin, where
$\mathbb{L}=1$, it is seen immediately from comparing~(\ref{bfjx})
and~(\ref{gruppis}) that
\begin{eqnarray}
\left.\mathcal{P}^x_{\Lambda=1,2,3}\right|_{q^u=0}&=&0\,,\qquad
\left.\mathcal{P}^x_{\Lambda=3+y}\right|_{q^u=0}\,=-\,r_1\,e_1\,\delta^x_y
\,.\label{Plam0}
\end{eqnarray}
The above property allows us to obtain at $q^u\,\equiv\, 0$ the analogue
of the $\SO(3)$ FI term, which in the model without hypermultiplets was
introduced by hand. Here, this term appears if the $\SO(3)$ part of the
isometry of the hypermultiplet manifold is gauged, i.e., $r_1=1$, and is
again proportional to the $\SO(3)$ charge $e_1$.

The extremum of $\mathcal{V}$ corresponds to the point
\begin{eqnarray}
\phi^{(0)}&=&\cases{y^a\equiv 0\,,\cr q^u\equiv 0\,,\cr S=S^{(0)}=\cot
\theta -{\rm i}\,\left|\frac{e_0\,\sqrt{1+2r_0^2}}
{\sqrt{3}\,r_1\,e_1\,\sin{(\theta)}}\right|\,, }\nonumber\\
\left.\mathcal{V}\right|_0 &=&
|\sqrt{3\,(1+2r_0^2)}\,r_1\,e_1\,e_0\,\sin{(\theta)}|>0\,. \label{vacuum}
\end{eqnarray}
We see that $r_1=0$ leads to a singular value for $S^{(0)}$, which
corresponds to the remarks in the previous models that one needs a FI
term. Thus we further restrict to $r_1=1$.

{}From the above analysis we may conclude that {\bf the potential has a
dS critical point which is placed at the origin of the quaternionic
manifold.} Another property of the critical point $\phi^{(0)}$ is that
\begin{equation}
L^\Lambda {\mathcal P}_\Lambda ^x = 0\,, 
\end{equation}
where we have used  (\ref{Plam0}) and that
$L^\Lambda\,\propto\,\{1/2,\,{\rm i}/2,\,0,\,0,\,0,\,0\}$ at $y^a=0$.\par

Let us now compute the mass matrix for the scalar fields. This matrix is
most easily expressed with respect to the fluctuation of the scalar
fields around the vacuum (\ref{vacuum}), using for the quaternionic
manifold the parametrization in terms of the $\tilde{y}^a$ coordinates.
The linear relation  between the fluctuations $\{\delta \tilde{y}^a\}$
and $\{\delta a_i,\,\delta b_1,\,\delta b_2,\,\delta h_1,\,\delta h_2\}$
around the origin of the quaternionic manifold can be derived from the
formal relations given in the appendix  and are
\begin{eqnarray}
\delta \tilde{y}_0&=& \frac{{\rm
i}}{2\,\sqrt{2}}\,(a_1-a_2)+\frac{1}{2}h_2\,,\qquad
 \delta \tilde{y}_1 = \frac{1}{2\,\sqrt{2}}\,(a_1+a_2)+\frac{{\rm i}}{2}\,h_1\,,\nonumber\\
 \delta \tilde{y}_2&=&\ft12\left(b_1+{\rm i}\,a_3\right)\,,\qquad
  \delta \tilde{y}_3 = \ft12\left(b_2+{\rm i}\,a_4\right).
\end{eqnarray}

We now split into two cases whether the $\SO(2,1)$ part of the isometry
group is gauged ($r_0=1$) or not ($r_0=0$).

\paragraph {$r_0=0$ :}
In this case, the scalar fluctuations in the special K{\"a}hler manifold and
those in the quaternionic manifold do not mix. The eigenvalues of the
mass matrix are
\begin{eqnarray}
r_0 = 0 && \cases{\left.\frac{\partial _\alpha \partial^ \beta
\mathcal{V}}{\mathcal{V}}\right|_{0} =\left(  \matrix{ 2 & 0 & 0 \cr
        0 & 0 &  0 \cr
        0 & 0 &  \unity _3 \cr  }\right) \,;\,\,\alpha,\,\beta\,\,\mbox{over complex scalars in vector multiplets}\cr
\left.\frac{\partial _a \partial^b\mathcal{V}}{\mathcal{V}}\right|_{0} =
\left(  \matrix{ 0 & 0 \cr
        0 & \frac{2}{3}\, \unity _3 \cr  }\right) \,;\,\,a,\,b\,\,\mbox{over the 4 complex hyperscalar fluctuations $\delta\tilde{y}^a$}}
\end{eqnarray}
Besides the complex Goldstone $\delta y_0$ (recall that $\langle
y_0\rangle=0$ is the only \textit{vev} which is not gauge invariant, in
particular it transforms under the action of the non-compact isometries
inside $\SO(2,1)$) we gain one more complex zero mode corresponding to
$\delta \tilde{y}_0$ which survives the BEH mechanism as zero mode of the
effective theory and is a singlet with respect to the residual gauge
group $\SO(2)\times \SO(3)$.

\paragraph{$r_0=1$ :}
In this case, the eigenstates of the mass matrices at $\phi^{(0)}$ are
mixed states between the Calabi--Vesentini scalars of the vector
multiplets and the corresponding ones in the hypermultiplets (here it is
clear that the alternative parametrization is most suitable). As it can
be easily checked from the Killing vectors (\ref{so21kil}) and their
analogous expressions in the quaternionic coordinates $\tilde{y}^a$, the
generator of $\SO(2)\subset \SO(2,1)$ defines in the tangent space of the
special K{\"a}hler and quaternionic-K{\"a}hler manifolds, at the points $y^a=0$
and $\tilde{y}^a=0$ respectively, the complex structures for $y^a$ and
$\tilde{y}^a$. Indeed, its infinitesimal action on the  fluctuations
$\delta y^a$ and $\delta\tilde{y}^a$ around the corresponding origins is
represented by  the Killing vector $k_3$ and has the linear form
\begin{eqnarray}
\SO(2):\,\,\cases{\delta y^a\rightarrow \delta y^a+\epsilon k_3^a(\delta
y)=\delta y^a+ {\rm i}e_0 \epsilon \delta y^a\,,\cr \delta
\tilde{y}^a\rightarrow\delta \tilde{y}^a+\epsilon  k_3^a(\delta
\tilde{y})=\delta\tilde{y}^a+  {\rm i}e_0 \epsilon\delta \tilde{y}^a}\,.
\end{eqnarray}
Since this transformation is part of the $\SO(2)\times\SO(3)$ residual
symmetry of our vacuum, we expect each term of the effective Lagrangian
to be invariant under its action. The effective mass term that we find
for the scalar fields can be expressed in terms of
 the variables $Z^\sigma=\delta y^a\pm \delta \tilde{y}^a$ in the form
\begin{eqnarray}
(m_S)^2\,\delta
S\delta\bar{S}+\sum_{\sigma=1}^8({m}_{\sigma})^2\,Z^\sigma\,\bar{Z}^{\sigma}\,,
\end{eqnarray}
which is manifestly invariant under the  residual $\SO(2)$ infinitesimal
transformation $Z^\sigma\rightarrow Z^\sigma+{\rm i}e_0\epsilon
\,Z^\sigma $. The generator of this $\SO(2)$ symmetry of the vacuum
defines therefore a complex structure for the scalar fields in the
effective theory deriving from $y^a$ and $\tilde{y}^a$. \par

The states $\delta S,\,Z^\sigma$ and their masses $m_S,\,m_\sigma$ in
units of $\left.\mathcal{V}\right|_0$ are listed in
table~\ref{tbl:masses}.
\begin{table}[ht]
  \caption{\it Mass square eigenvalues for $r_0=1$ in units of
  $\left.\mathcal{V}\right|_0$ and corresponding scalar fluctuations, indicating the number of (complex)
  scalars with the corresponding value and representation with respect to the residual gauge group $\U(1)\times\SO(3)$.}\label{tbl:masses}
\begin{center}
\begin{tabular}{|c|c|c|c|}\hline\hline\rule[-1mm]{0mm}{6mm}
 {$m^2/\left.\mathcal{V}\right|_0$} & number & eigenstate &$\U(1)\times\SO(3)$ rep.\\[2mm]\hline
 \rule[-1mm]{0mm}{7mm}
$2$&1&$ \delta S  $  & {\bf $(1,1)$} \\[3mm] \hline \rule[-1mm]{0mm}{7mm}
 $0$&1&$ \delta y_0+\delta\tilde{y}_0 $ & \\[3mm] \hline \rule[-1mm]{0mm}{7mm}
 $ \frac{2}{3}$  & 1 &    $ \delta y_0-\delta\tilde{y}_0$ & {\bf $(1_{\mathbb{C}},1)$} \\[3mm] \hline \rule[-1mm]{0mm}{7mm}
$\frac{4 }{3}$&3& $ \delta \vec y+\delta\vec {\tilde{y}}$& {\bf $ (1_{\mathbb{C}},3)$} \\[3mm] \hline \rule[-1mm]{0mm}{7mm}
$0$&3& $ \delta \vec y-\delta\vec {\tilde{y}}$& {\bf $(1_{\mathbb{C}},3)$} \\[3mm]\hline \hline
\end{tabular}
\end{center}
\end{table}
The complex zero mode $\delta y_0+\delta\tilde{y}_0$ is the complex
Goldstone boson associated with the broken non-compact $\SO(2,1)$
transformations and therefore, its  real and imaginary components are
`eaten' by the vector fields $A_\mu^1,\,A_\mu^2$ according to the BEH
mechanism analogous to the one described in section~\ref{so21so3}.
Besides this Goldstone zero mode, we have 3 more complex zero modes
$\delta \vec y-\delta\vec{\tilde{y}}$, which survive the BEH mechanism
and whose real and imaginary components transform in the ${\bf (2,3)}$ of
the residual gauge group $\SO(2)\times \SO(3)$. \bigskip

We thus find again that in both the cases {\bf  the potential admits a
stable dS vacuum}, though there are now valleys in the scalar potential,
which are reminiscent of the valleys that were present also in the rigid
model with hypermultiplets of~\cite{Kallosh:2001tm}. One may wonder
whether quantum effects could lead to a sliding of the scalar
\textit{vev}. An indication of this could be given by the fact whether
the valley `narrows' or `broadens' when one follows it. This is shown by
the values of the masses in other vacua than~(\ref{vacuum}). We did a
perturbative analysis going to neighbouring points in the valley for the
$r_0=0$ case and found that the masses do not change to the order that we
considered.

\section{Summary and conclusions: embedding into superstring theory?}\label{ss:summOutlook}
In the present paper, we have shown that within the framework of standard
matter-coupled $\mathcal{N}=2$ supergravity, theories admitting
\textit{stable de Sitter vacua} do exist. We have shown two models with
only positive mass fields and one which has also a flat valley in the
moduli space of hypermultiplets, showing some similarities to a
corresponding model in rigid supersymmetry~\cite{Kallosh:2001tm}. We have
emphasized that the catch to obtain such a positive result is the use of
three equally essential ingredients, namely:
\begin{enumerate}
  \item non-compact, non-Abelian gaugings;
  \item de Roo--Wagemans angles corresponding to symplectic rotations
  of one simple gauge group factor with respect to another; and
  \item the presence of Abelian or non-Abelian Fayet--Iliopoulos terms in
  the case of pure vector multiplet theories, alternatively coupling to
  hyper multiplets in such a way that one obtains an effective
  Fayet--Iliopoulos term produced by the hyper \textit{vev}s.
\end{enumerate}
We have illustrated our positive result by analysing three models of
increasing complexity by means of which we were able to show the role of
the three ingredients in obtaining the final outcome. Essentially we can
say that:
\begin{enumerate}
  \item The use of non-compact gaugings contributes, into the formula
  of the scalar potential the positive definite term
\begin{equation}
  k^{\alpha }_\Lambda \, k^{{\bar \beta}}_\Sigma \, g_{\alpha {\bar \beta}} \,
 {\bar L}^\Lambda \, L^\Sigma\,,
\label{cupi}
\end{equation}
which cannot be absorbed into a superpotential.
  \item The de Roo--Wagemans angles  are essential in
  order to introduce a non-trivial \textit{dilaton} dependence of the
  scalar potential and hence to allow for extrema. Without such
  angles, the scalar potential would simply be
\begin{equation}
  V_{\rm scalar} = \frac{1}{{\rm Im} \, S} \, \times \, \mbox{function $V^\prime$ of all other scalars.}
\label{nobono}
\end{equation}
\item Finally, the Fayet--Iliopoulos term or the coupling to hypers
contributes the source term that yields a finite value to the
\textit{vev} of the dilaton, once the de Roo--Wagemans angles are
included.
\end{enumerate}
The combination of these ingredients avoids the negative conclusions
reached in (\ref{eigenvalue-2}). Important in that respect is that in the
vacua the gravitino shifts [$S_{ij}$ in~(\ref{susyFermTrans})], which are
zero due to the setting where  $L^\Lambda$ is orthogonal to ${\cal
P}^x_\Lambda =0$.  This removes the eigenvector that would generalize the
one found in~\cite{Cremmer:1985hj} and further discussed in
section~\ref{ss:AbelianDS}. This is the technical understanding of the
mechanisms producing the de Sitter stable vacuum. \par

The next two questions which are intimately related are:
\begin{description}
  \item[a] What is the physical relevance of the three ingredients
  quoted above?
  \item[b] How can our result be lifted to higher $\mathcal{N}$
  supergravities and be embedded into superstring theory or
  M-theory?
\end{description}
\par
To provide a first provisional answer to question $[a]$, we emphasize
that gauged supergravities in $D=p+2$ dimensions emerge as the
\textit{near-brane} description of light bulk mode interactions in the
geometry produced by $p$-brane configurations. This is fully understood
for compact gaugings like the $\SO(6)$ gauging of $\mathcal{N}=8$
supergravity in $D=5$, which emerges as the near-horizon description of
the $D3$-brane, or for the $\SO(8)$ gauging  of $\mathcal{N}=8$
supergravity in $D=4$, which is associated with the near horizon
$M2$-brane. This relation is much more poorly understood for non-compact
gaugings, yet it is clear that also there one should be able to trace
back the gauging to suitable brane constructions. In view of this, the de
Roo--Wagemans rotation, which is a symplectic rotation turning part of
the electric fields into magnetic ones, needs to be interpreted in terms
of its action on candidate branes participating in the construction. Such
an analysis is postponed to future publications and investigations. It
is, however, worth mentioning here that the de Roo--Wagemans angles were
originally introduced in the context of $\mathcal{N}=4$ supergravity and
therefore do exist also in higher $\mathcal{N}$ theories. Indeed it has
begun to be appreciated only recently that the classification of
\textit{gauged} supergravities can be extended if the \textit{electric
group} is modified, which is just what the de Roo--Wagemans rotation
does. In particular, the recent results on new $\mathcal{N}=8$
gaugings~\cite{Andrianopoli:2002mf,Hull:2002cv} pertain to such a
scenario. In \cite{Cordaro:1998tx}, the exhaustive classification of
$\mathcal{N}=8$ gaugings was obtained under the dogma that the electric
group should be
\begin{equation}
  G_{\rm electric} = \mathrm{SL\left (8,\mathbb{R}\right )} \subset
  \mathrm{E_{7(7)}}\,,
\label{dogma}
\end{equation}
and it appears that within such classification no stable de Sitter vacuum
is contained. If the dogma (\ref{dogma}) is removed, then new gaugings,
as~\cite{Andrianopoli:2002mf,Hull:2002cv} have proven, are possible and
the question is  reopened whether stable de Sitter vacua could be present.
\par
This brings the discussion to question $[b]$, namely whether the
successful $\mathcal{N}=2$ models can be lifted to higher $\mathcal{N}$
and possibly interpreted within string theory. In this respect, the main
observation is that the third most complex model, that including the
hypermultiplets, is not just randomly chosen but it is a very specific
one with a quite inspiring motherhood. To see this consider
$\mathcal{N}=4$ supergravity coupled to $n=n_1 + n_2$ vector multiplets.
The scalar manifold is
\begin{equation}
  M_{\rm scalar}^{{\cal N}=4} = \frac{\SU(1,1)}{\U(1)} \, \times \, \frac{\SO(6,n)}{\SO(6) \times \SO(n)
  }\,.
\label{n4scala}
\end{equation}
Such a theory can be truncated to $\mathcal{N}=2$ and a useful and
consistent way to do it is by modding with respect to a discrete subgroup
of the holonomy group $H_{\rm hol} =\SO(6) \times \SO(n)$. For instance,
if $\alpha$ denotes a square root of the identity ($\alpha^2=1$) one can
embed a $\mathbb{Z}_2$ group into the holonomy group in the following
way:
\begin{equation}
 \mathbb{Z}_2 \ni \alpha \, \hookrightarrow \, \left(\begin{array}{c|c}
   {\bf 1}_{2 \times 2} & 0 \\
  \hline
   0 & \alpha \,  {\bf 1}_{4 \times 4} \
 \end{array} \right)  \,
 \otimes \, \left( \begin{array}{c|c}
   {\bf 1}_{n_1 \times n_1} & 0 \\
   \hline
   0 & \alpha \, {\bf 1}_{n_4 \times n_4} \
 \end{array}\right) \, \in \, H_{\rm hol}\,.
\label{aletto!}
\end{equation}
There are two gravitinos that survive such  a $\mathbb{Z}_2$ projection
and the scalars that are $\mathbb{Z}_2$ singlets span the manifold
\begin{equation}
  \frac{\SU(1,1)}{\U(1)} \, \times \, \frac{\SO(2,n_1)}{\SO(2) \times \SO(n_1) }
  \times \, \frac{\SO(4,n_2)}{\SO(4) \times \SO(n_2) }\,.
\label{n1+n2}
\end{equation}
For $n_1=4$ and $n_2=2$, the above manifold just corresponds to the
$\mathcal{N}=2$ model with hypers studied in the present paper. Since
$2+4=6$, this means that our successful model can be embedded into an
$\mathcal{N}=4$ supergravity with six vector multiplets and based on the
scalar manifold
\begin{equation}
   \mathcal{ST}[6,6] = \frac{\SU(1,1)}{\U(1)} \, \times \, \frac{\SO(6,6)}{\SO(6) \times \SO(6)
   }\,.
\label{mt66}
\end{equation}
As for the gauge group we can just choose
\begin{equation}
  G_{\rm gauge}^{{\cal N}=4} = \SO(2,2) \times \SO(4) \sim \left[  \SO(2,1)\times \SO(3)\right]^2\,,
\label{n4gaug}
\end{equation}
which has 12 generators in the fundamental of $\SO(6,6)$ and contains two
copies of the $\mathcal{N}=2$ gauge group. Indeed the latter is just the
diagonal subgroup. What remains to be proven and is left to a future
publication is that the $\mathcal{N}=4$ potential is extremum at a zero
value of the additional  scalars that are not $\mathbb{Z}_2$ singlets. If
that is true, the embedding of our model into $\mathcal{N}=4$ is perfect.
\par
On the other hand, the scalar manifold (\ref{mt66}) is just the standard
moduli space for the toroidal compactification of type IIA string theory
on a $T^6$ torus. Indeed, $\mathcal{ST}[6,6] \subset E_{7(7)}/\SU(8)$ is
just the submanifold of Neveu--Schwarz scalars in an $\mathcal{N}=8$
theory.
\par
It follows from these observations that the prospects to reinterpret our
stable de Sitter vacuum as a vacuum in a brane construction and within
the framework of superstring theory are, at first sight, quite promising.

\medskip
\section*{Acknowledgments}

\noindent We are grateful to Renata Kallosh for several discussions,
which, in particular, gave rise to the calculations leading to the mass
matrices. Furthermore, we thank Gianguido Dall'Agata for contributions to
initial stages of our investigations, and G. Gibbons for interesting
discussions.

\newpage
\appendix
\section{Quaternionic geometry items}
In the main text, the essential ingredients of the quaternionic part of
the scalar potential, namely the vielbein, the quaternionic metric
$h_{uv}$, the triholomorphic moment maps $ \mathcal{P}^x_\Lambda$ and the
triholomorphic Killing vectors $k^u_\Lambda$ have been defined
through~(\ref{defEfromcoset}), (\ref{quatmetric}),  (\ref{Plam}) and
(\ref{trikillo}). In order to make such definitions explicit functions of
the fields and workable by the reader one just needs to specify the
explicit form of the involved matrices, namely the coset representative
$\mathbb{L}$, that defines the vielbeins. The latter determine the
metric, which we will give explicitly and the complex structures $J^x$.
For the gauging, we also need the group generators $t_\Lambda$. This is
what we do in this appendix. For the sake of explicit calculations, the
solvable parametrization is much simpler than any other parametrization
and for this reason we use it, and make it explicit in
section~\ref{app:solv}. In section~\ref{app:CV} of this appendix, we
discuss the relation between the solvable coordinates and the
Calabi--Vesentini coordinates for the same manifold.

\subsection{The solvable parametrization}\label{app:solv}

\paragraph{Coset representative, vielbein and metric.}
The coset representative $\mathbb{L} $ in solvable coordinates is given
by a $6 \times 6$ matrix which, both for convenience and for further
manipulations, can be written in block form as follows:
\begin{equation}
  \mathbb{L}_{\rm solv} = \left( \begin{array}{c||c}
    \underbrace{A_{\rm solv}}_{4 \times 4} & \underbrace{B_{\rm solv}}_{4 \times 2 }\\
    \hline
    \hline
    \underbrace{C_{\rm solv}}_{2\times 4} & \underbrace{D_{\rm solv}}_{2\times2} \
  \end{array} \right)
\label{bloccus}
\end{equation}
The explicit forms of the blocks can be displayed and are polynomial in
the nilpotent fields $a_i$, $b_i$, while their dependence on the Cartan
fields $h_i$ is via simple exponentials, namely
\begin{eqnarray}
  A_{\rm solv}&=&\pmatrix{  1 & 0&{a_4} + \sqrt{2}\,{a_1}\,{b_2} &
 {a_3} + \sqrt{2}\,{a_1}\,{b_1} \cr
  0 & 1 &{b_2} &    {b_1}\cr
 -a_4{\rme^{{-h_1}}}   & -b_2\rme^{-h_2}  &
\cosh h_1+c&\frac{1}{\sqrt{2}}\left( a_1 \rme^{h_2}-
a_2\rme^{-h_2}\right) -d \cr
    -a_3\rme^{-h_1}  & -b_1\rme^{-h_2}  &
      \frac{1}{\sqrt{2}} \rme^{-h_1}\left( -a_1 + a_2\right)
   & \cosh h_2-\ft12b^2\rme^{-h_2}    }\,,
   \nonumber\\[2mm]
B_{\rm solv} & = &  \pmatrix{
 \rme^{-h_1}a_4 & \rme^{-h_2}b_2 \cr
 \rme^{-h_1} a_3 & \rme^{-h_2} b_1\cr
\sinh h_1 -c   & \frac{1}{\sqrt{2}}\left( a_1 \rme^{h_2}+
a_2\rme^{-h_2}\right) +d \cr
 \frac{1}{\sqrt{2}}\rme^{-h_1}\left( a_1 - a_2\right)\qquad
    & \sinh h_2 +\ft12 b^2 \rme^{-h_2}   }\,,
    \nonumber\\[2mm]
C^T_{\rm solv} & = &  \pmatrix{a_4 + \sqrt{2}a_1b_2 & b_2 \cr
 a_3 + \sqrt{2}a_1b_1 & b_1 \cr
 \sinh h_1 +c  & \frac{1}{\sqrt{2}}\rme^{-h_1}\left( a_1+ a_2\right) \cr
 \frac1 {\sqrt{2}}\left( a_1\rme^{h_2} -  a_2\rme^{-h_2}\right)-d\quad &
  \sinh h_2    - \ft12 b^2 \rme^{-h_2} \cr
     }\,,\nonumber\\[2mm]
D_{\rm solv} & = & \pmatrix{\cosh h_1 -c  &  \frac1 {\sqrt{2}}\left(
a_1\rme^{h_2} +  a_2\rme^{-h_2}\right)+d\cr
 -\frac{1}{\sqrt{2}}\rme^{-h_1}\left( a_1 + a_2\right)  \qquad &\cosh h_2+\ft12 b^2\rme^{-h_2} } \,,
\label{BCDblocks}
\end{eqnarray}
with
\begin{equation}
b^2\equiv b_1{}^2+b_2{}^2\,,\qquad
 c\equiv   \ft12\rme^{-h_1}\left(  2a_1a_2
       - a_3{}^2 - a_4{}^2 \right)\,,\qquad
d\equiv     \rme^{-h_2} ( a_3b_1 + a_4b_2   +
 \ft{1}{\sqrt{2}}a_1 \,b^2)\,.
 \label{defcd}
\end{equation}
These lead by~(\ref{defEfromcoset}) to the vielbein
\begin{equation}
V^{mt}=  \pmatrix{\rme^{-h_1}\left( \rmd a_4+\sqrt{2}\,b_2\,\rmd
a_1\right) &\rme^{-h_2}\,\rmd b_2\cr
 \rme^{-h_1}\left( \rmd a_3+\sqrt{2}\,b_1\, \rmd
a_1\right)&\rme^{-h_2}\,\rmd b_1\cr
 \rmd h_1 & \frac{1}{\sqrt{2}}\rme^{-h_2}\left(\rme^{h_1} \rmd a_1+\rme^{-h_1}A_2\right)\cr
\frac{1}{\sqrt{2}}\rme^{-h_2}\left(\rme^{h_1} \rmd
a_1+\rme^{-h_1}A_2\right)&\rmd h_2}\,,
 \label{Vmtexpl}
\end{equation}
with
\begin{equation}
A_2=\rmd   a_2+
  b^2\rmd a_1+\sqrt{2}(b_1\rmd a_3 +b_2 \rmd a_4)\,.
 \label{A1A2}
\end{equation}

The vielbein defines the metric by~(\ref{quatmetric}). We obtain
\begin{eqnarray}
 2 \rmd s^2&=&\rme^{-2(h_1+h_2)}a_{ij}\rmd a_i\rmd a_j  +\rme^{-2h_2}\left[
  (\rmd b_1)^2+(\rmd b_2)^2\right]+(\rmd h_1)^2+(\rmd h_2)^2
  \,,\nonumber\\[2mm]
a_{ij}&=&\pmatrix{\left( \rme^{2\,h_2} + b^2  \right)^2
   &  b^2 & \sqrt{2}\,b_1
     \left( \rme^{2\,h_2} + b^2  \right) &\sqrt{2}\, b_2\,
     \left( \rme^{2\,h_2} + b^2  \right) \cr
      b^2   & 1 &  \sqrt{2}\,   b_1& \sqrt{2}\,b_2 \cr
    \sqrt{2}\,b_1\,
     \left( \rme^{2\,h_2} + b^2  \right) & \sqrt{2}\,b_1
    & \rme^{2\,h_2} + 2\,{b_1}^2     & 2\,b_1\,b_2     \cr
   \sqrt{2}\, b_2\left( \rme^{2\,h_2} + b^2  \right)  &\sqrt{2}\, b_2
    & 2\,b_1\,b_2 & \rme^{2\,h_2} + 2\,{b_2}^2
   }\,.
 \label{huv}
\end{eqnarray}

\paragraph{The group generators.}
The generators of $\SO(2,1)\times \SO(3) \subset \SO(2,4)$ are the $6
\times 6 $ matrices already spelt out in (\ref{Tlamdefi}). Indeed, the
quaternionic manifold is just a copy of the submanifold $\SO(2,4)/\SO(2)
\times \SO(4)$ of $\mathcal{ST}[2,4]$ and therefore the generators of the
isometry algebra are the same. One has just to be careful with the fact
that for the use in the vector multiplet sector the $6$-dimensional
representation of $SO(2,4)$ was written in the basis where the $2 \times
2 $ block is the first while the $4 \times 4$ block is the last. For the
use in the quaternionic case, the same matrices have to be written in a
basis where the $4 \times 4$ block is instead the first. Hence it
suffices to take the matrices (\ref{Tlamdefi}) and do the permutation of
axis:
\begin{equation}
  \left\{ 1,2,3,4,5,6\right\} \Rightarrow \left\{ 6,5,4,3,2,1\right\}.
\label{permutazia}
\end{equation}
Explicitly, we obtain
\begin{equation}
  \begin{array}{ccccccc}
    t_1 & = & e_0\left( \matrix{
   0 & 0 & 0 & 0 & 0 & 0 \cr 0 & 0 & 0 & 0 & 0 & 0 \cr
   0 & 0 & 0 & 0 & 0 & 0 \cr 0 & 0 & 0 & 0 &
    -1 & 0 \cr 0 & 0 & 0 &
    -1 & 0 & 0 \cr 0 & 0 & 0 & 0 & 0 & 0 \cr  }\right),  & \qquad
    & t_2 & = & e_0\left(\matrix{
   0 & 0 & 0 & 0 & 0 & 0 \cr 0 & 0 & 0 & 0 & 0 & 0 \cr
   0 & 0 & 0 & 0 & 0 & 0 \cr 0 & 0 & 0 & 0 & 0 & 1 \cr
   0 & 0 & 0 & 0 & 0 & 0 \cr 0 & 0 & 0 & 1 & 0 & 0 \cr
    }\right),   \\
    \null & \null & \null & \null & \null & \null\null \\
    t_3 & = & e_0\left( \matrix{
   0 & 0 & 0 & 0 & 0 & 0 \cr 0 & 0 & 0 & 0 & 0 & 0 \cr
   0 & 0 & 0 & 0 & 0 & 0 \cr 0 & 0 & 0 & 0 & 0 & 0 \cr
   0 & 0 & 0 & 0 & 0 & 1 \cr 0 & 0 & 0 & 0 &
    -1 & 0 \cr  }\right),  & \qquad & t_4 & = & e_1\left( \matrix{ 0 & 1 & 0 & 0 & 0 & 0 \cr
    -1 & 0 & 0 & 0 & 0 & 0 \cr 0 & 0 & 0 & 0 & 0 &
   0 \cr 0 & 0 & 0 & 0 & 0 & 0 \cr 0 & 0 & 0 & 0 & 0 &
   0 \cr 0 & 0 & 0 & 0 & 0 & 0 \cr  }\right),  \\
   \null & \null & \null & \null & \null & \null\null \\
    t_5 & = & e_1\left( \matrix{ 0 & 0 &
    -1 & 0 & 0 & 0 \cr 0 & 0 & 0 & 0 & 0 & 0 \cr 1 &
   0 & 0 & 0 & 0 & 0 \cr 0 & 0 & 0 & 0 & 0 & 0 \cr 0 &
   0 & 0 & 0 & 0 & 0 \cr 0 & 0 & 0 & 0 & 0 & 0 \cr  }\right),  & \qquad &
   t_6 & = & e_1 \left(
   \matrix{
   0 & 0 & 0 & 0 & 0 & 0 \cr 0 & 0 & 1 & 0 & 0 & 0 \cr
   0 &
   -1 & 0 & 0 & 0 & 0 \cr 0 & 0 & 0 & 0 & 0 & 0 \cr
   0 & 0 & 0 & 0 & 0 & 0 \cr 0 & 0 & 0 & 0 & 0 & 0 \cr
    }\right).
  \end{array}
\label{gruppis}
\end{equation}
\par

Rather than giving the other relevant items in terms of the solvable
coordinates it is more interesting to explore the relation between the
latter and the Calabi--Vesentini coordinates for the same manifold which
we have already used in the vector multiplet sector.
\subsection{Transformation from the solvable to the Calabi--Vesentini
coordinates} \label{app:CV}

The coset manifold
\begin{equation}
  M_{2,4} \equiv \frac{\SO(2,4)}{\SO(2) \times \SO(4)}
\label{so24man}
\end{equation}
is remarkable in that it can be alternatively seen as a complex manifold
of the series
\begin{equation}
  \frac{\SO(2,n)}{\SO(2) \times \SO(n)}\,,
\label{2nseries}
\end{equation}
or a quaternionic manifold of the series
\begin{equation}
  \frac{\SO(4,m)}{\SO(4) \times \SO(m)}\,.
\label{4mseries}
\end{equation}
The double interpretation implies that although it is quaternionic, yet
it admits a description in terms of the Calabi--Vesentini complex
coordinates already used in the case of vector multiplets. In this
section, we elaborate the coordinate transformation from the solvable
basis to the Calabi--Vesentini basis.
\par

Our starting point is provided by equations (C.1)--(C.4) of
\cite{Andrianopoli:1997cm}. It follows from there that if
\begin{equation}
  \mathbb{L}_{\rm CV} = \left( \begin{array}{c||c}
    \underbrace{A_{\rm CV}}_{4 \times 4} & \underbrace{B_{\rm CV}}_{4 \times 2 }\\
    \hline
    \hline
    \underbrace{C_{\rm CV}}_{2\times 4} & \underbrace{D_{\rm CV}}_{2\times2} \
  \end{array} \right)
\label{bloccusCV}
\end{equation}
is the coset representative in the CV basis, the upper part of the
symplectic section
\begin{equation}
   {X}^\Lambda  =  \left( \begin{array}{c}
  \ft 1 2 \, \left( 1+{  y}^2\right)  \\
  \ft 12 \, {\rm i} \, (1-{  y}^2) \\
  {  y}^\alphaa
\end{array}\right)
\label{tilsec}
\end{equation}
is related to the matrix  blocks in the following way. Let
\begin{equation}
  \mathbb{S}=\left( \begin{array}{c||c}
    \underbrace{\unity }_{4 \times 4} & \underbrace{0}_{4 \times 2 }\\
    \hline
    \hline
    \underbrace{0}_{2\times 4} & \underbrace{\mathcal{T}}_{2 \times 2} \
  \end{array} \right),
\qquad \mbox{where}\qquad
  \mathcal{T}= \frac{1}{\sqrt{2}}\left( \begin{array}{cc}
    {\rm i} & -{\rm i} \\
    1 & 1 \
  \end{array} \right).
\label{taumatra}
\end{equation}
Defining
\begin{equation}
  \widehat{\mathbb{L}}_{\rm CV}= \mathbb{S}^{-1} \, \mathbb{L}_{\rm CV} \, \mathbb{S}
  =\left( \begin{array}{c||c}
    \widehat{A}_{\rm CV} & \widehat{B}_{\rm CV} \\
    \hline
    \hline
    \widehat{C}_{\rm CV} & \widehat{D}_{\rm CV} \
  \end{array}\right),
\label{lupicanti}
\end{equation}
one has
\begin{equation}
\widehat{B}_{\rm CV}  =  \exp[\mathcal{K}_2/2] \, \left(
\begin{array}{c}
  y^a \\
  {\bar y}^a
\end{array}\right) \,,\qquad
\widehat{D}_{\rm CV} =  \frac{1}{\sqrt{2}} \, \exp[\mathcal{K}_2/2] \,
\left(
\begin{array}{cc}
  1 & y^2 \\
  {\bar y}^2 & 1
\end{array} \right ),
\label{Dhat}
\end{equation}
where
$y^2  = y^a y^a $
 and ${\bar y}^2   =  {\bar y}^a {\bar y}^a$
and $\mathcal{K}_2$ is the K{\"a}hler potential as given in
(\ref{kalermetr}). Consider next the following $2\times 4$ matrix:
\begin{equation}
  \mathbb{IP}  \equiv \widehat{B}  \, \cdot \,
  \widehat{D}^{-1} \equiv \left(
    Y^a \, , \,  {\bar Y}^a\right).
\label{Ydefi}
\end{equation}
By explicit calculation we obtain
\begin{equation}
  Y^a = \sqrt{2} \frac{1}{1-|y^2|^2} \, \left( y^a - y^2 \, {\bar
  y}^a \right).
\label{BigYsmally}
\end{equation}
This relation can be inverted by means of the following formula:
\begin{equation}
  y^a = \frac{1}{\sqrt{2}} \left(\, Y^a + \, t \, {\bar Y}^a \right),\qquad \mbox{with}
\qquad   t  =  \frac {\left( 1-Y\, \cdot \, \overline{Y}\right)
-\sqrt{\left( 1-Y\, \cdot \,
  \overline{Y}\right)^2-|Y \, \cdot \, Y |^2}}{Y\, \cdot \,
  \overline{Y}}\,.
\label{inverto}
\end{equation}
Why do we consider the matrix $\mathbb{IP}$ defined in (\ref{Ydefi})? The
reason is simple. By construction it is a projective invariant that is
independent from the choice of the coset representative out of which it
is constructed. It depends only on the equivalence class, namely on the
point of the coset manifold. Indeed, since the subgroup $\SO(4) \times
\SO(2)$ is block diagonal, under a transformation
\begin{equation}
  \mathbb{L} \mapsto \mathbb{L}^\prime =\mathbb{L} \, \left( \begin{array}{c||c}
    H_{4} & 0 \\
    \hline\hline
    0 & H_{2}
  \end{array} \right),
\label{pidoccio}
\end{equation}
the matrix $ \mathbb{IP}$ remains invariant. Hence, although  the coset
representatives calculated in the solvable and in the CV parametrizations
are different choices of representatives in the same equivalence classes,
we can safely identify
\begin{equation}
  \mathbb{IP}_{\rm solv} = \mathbb{IP}_{\rm CV}\,.
\label{solv_CV}
\end{equation}
Equation (\ref{solv_CV}) combined with (\ref{inverto}) provides the
desired coordinate transformation expressing the Calabi--Vesentini
coordinates in terms of the solvable ones. It suffices to set
\begin{equation}
  y^a =\frac{1}{\sqrt{2}} \left(\, Y^a_{\rm solv} + \, t_{\rm solv} \, {\bar Y}^a_{\rm solv}
  \right),
\label{picchio}
\end{equation}
where in the r.h.s. the $Y^a$ and   $t(Y)$ are calculated from the
solvable coset representative.
\par

It turns out that the $Y^a$ are lengthy polynomial functions of the
solvable parameters and that the invariant $t$ is very much complicated
although completely explicit. So, as it might be expected, the
transformation
\begin{equation}
  y^a = y^{a}(a_i,b_1,b_2,h_1,h_2)
\label{ysolvato}
\end{equation}
is highly nonlinear and quite involved. Yet, near the origin of the coset
manifold, namely for very small fields, the transformation
(\ref{ysolvato}) linearizes and becomes fairly simple
\begin{equation}
  y^a \, \simeq \frac{1}{2}\left(
  \begin{array}{c}
    {\rm i} \,{a_4} + {b_2}    \\
    {\rm i} \,{a_3} + {b_1}    \\
    \frac1{\sqrt{2}}(a_1 + a_2) +{\rm i} h_1     \\
   \frac{\rmi}{\sqrt{2}} (a_1- a_2) +h_2
      \end{array}
      \right).
\label{linear}
\end{equation}
These $y^a$ are the $\tilde y^a$ in the main text, giving the alternative
parametrization of the quaternionic-K{\"a}hler manifold.
\section{Indices}
\begin{table}[ht]
  \caption{\it Indices in this paper, and their ranges, where $\dim SK$ stands for the (complex) dimension of the
  special K{\"a}hler manifold, and $\dim QK$ for the (quaternionic) dimension of the quaternionic-K{\"a}hler manifold.}\label{tbl:indices}
\begin{center}
  \begin{tabular}{|l|l|}
\hline\hline
 index and range & meaning \\
\hline
 $\mu =0,\ldots ,3$ & spacetime \\
 $i=1,2$ & supersymmetry extension \\
 $\Lambda =1,\ldots ,\dim SK +1$ & gauge group \\
 $x=1,2,3$ & $\SO(3)$ or $\SO(2,1)$ \\
 $\alpha,\bar \alpha  =1,\ldots ,\dim SK$ & complex scalars in $SK$\\
 $a=0,\ldots ,n-1$ & complex scalars in $\frac{\SO(2,n)}{\SO(2)\times \SO(n)}$ \\
 & (sometimes split in $y^0$ and $\vec y$)\\
 $A=1,\ldots ,6$ & $\SO(4,2)$ fundamental representation \\
 $u=1,\ldots ,4\dim QK$ & real scalars in $QK$ \\
 $t=1,\ldots ,\dim QK$  & quaternions \\
 $m=1,\ldots ,4$   & quaternionic components \\
 $i=1,\ldots ,4$ & Peccei-Quinn scalars in the solvable  \\
                  & parametrization of $QK$ (in
                  section~\ref{ss:withhyper} only)\\
$\sigma=1,\ldots ,8$ & complex scalars $Z^\sigma=\delta y^a\pm \delta
\tilde{y}^a$ in the  \\& effective theory around the dS vacuum\\&  (in
                  section~\ref{ss:withhyper} only)\\
\hline\hline
\end{tabular}
\end{center}
\end{table}
Table~\ref{tbl:indices} shows the indices used, their range and meaning.

\providecommand{\href}[2]{#2}\begingroup\raggedright\endgroup

\end{document}